\documentclass[11pt]{article}

\usepackage[utf8]{inputenc}
\usepackage[T1]{fontenc}
\usepackage[margin=1in]{geometry}
\usepackage{graphicx}
\usepackage{amsmath}
\usepackage{amssymb}
\usepackage{textcomp}
\usepackage{booktabs}
\usepackage{tabularx}
\usepackage{caption}
\usepackage[hidelinks,colorlinks=true,linkcolor=blue,citecolor=blue,urlcolor=blue]{hyperref}

\title{\Large\bfseries AI Hardware Accelerators for Large Language Models: Architectures and the Memory Wall}

\author{
  Siddharth Patel\thanks{Corresponding author. ORCID: 0009-0009-5384-0341. Email: \texttt{sp839@snu.edu.in}}
  \quad
  Rohit Singh\thanks{ORCID: 0000-0002-0789-337X. Email: \texttt{rohit.singh@snu.edu.in}}
  \\[6pt]
  \normalsize Department of Electronics Engineering \\
  \normalsize Shiv Nadar Institution of Eminence, Delhi-NCR, India
}
\date{}

\begin{document}

\maketitle

\begin{abstract}
\noindent Large language models (LLMs) place unprecedented and still-growing demands on the hardware that trains and serves them. This review surveys the full landscape of AI hardware accelerators for LLMs---general-purpose GPUs; custom ASICs such as TPUs, Trainium, Groq, and Cerebras; reconfigurable FPGAs; processing-in-memory and near-memory architectures; and emerging neuromorphic and photonic approaches---across cloud and edge deployment. Using the transformer's computational structure and roofline analysis as a common framework, we show that the decisive constraint on LLM acceleration is not arithmetic but memory: the autoregressive decode phase is bandwidth-bound, the key--value cache can rival the model weights in size, and data movement dominates energy. Comparing platforms on compute, memory, energy, programmability, and scalability, we find that no single architecture is optimal across workloads: GPUs remain the flexible default and the workhorse of training; domain-specific ASICs win at scale for stable, high-volume workloads; processing-in-memory is the most promising near-term response to the memory wall, entering systems as a heterogeneous complement; and neuromorphic and photonic computing, while promising, are not yet production-ready at frontier scale. Future progress depends on hardware--algorithm co-design and heterogeneous, memory-centric systems: for large language models, the memory system has become the computer.

\vspace{6pt}
\end{abstract}

\section{Introduction}
\subsection{The Large Language Model Revolution}
The introduction of the Transformer architecture marked a decisive break from the recurrent and convolutional models that preceded it, replacing sequential recurrence with a fully parallelizable self-attention mechanism and thereby unlocking training at unprecedented scale \cite{vaswani2017attention}. This architectural shift catalyzed a rapid succession of increasingly capable generative models, from GPT-2 and the 175-billion-parameter GPT-3 \cite{brown2020language} to GPT-4 \cite{openai2023gpt4}, the open-weight LLaMA family \cite{touvron2023llama,touvron2023llama2}, and a competitive landscape of contemporary systems including Gemini and Claude. Over roughly five years, the parameter counts of frontier models grew by more than three orders of magnitude---from the hundreds of millions typical of early Transformer encoders to figures reported to approach or exceed one trillion for the largest contemporary systems, although exact specifications for proprietary models remain undisclosed. This scaling has been accompanied by a broadening of application domains spanning conversational assistants, software synthesis, scientific discovery, and autonomous agents, each reinforcing the demand for ever-larger models and, by extension, ever-greater computational capacity.

\begin{figure}[htbp]
\centering
\includegraphics[width=\linewidth]{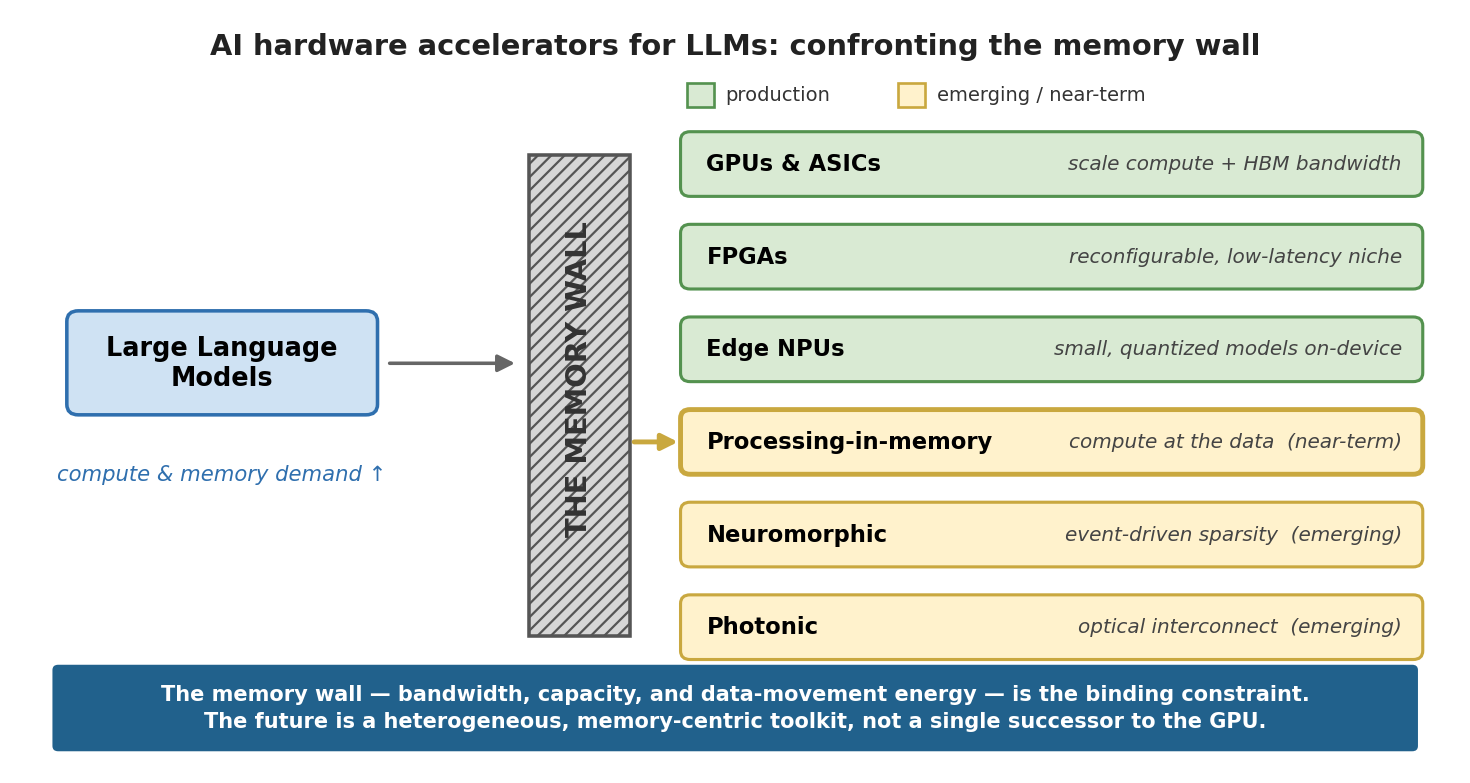}
\caption{Overview of the survey. Large language models press against the \emph{memory wall}---the bandwidth, capacity, and data-movement energy of the memory system---which this review identifies as the binding constraint on their acceleration. The six accelerator families examined (general-purpose GPUs and domain-specific ASICs, reconfigurable FPGAs, and edge NPUs in production today; processing-in-memory, neuromorphic, and photonic devices as emerging or near-term complements) are organized by maturity, and the central argument is that the field is converging on a heterogeneous, memory-centric toolkit rather than a single successor to the GPU.}
\label{fig:overview}
\end{figure}

\subsection{The Hardware and Energy Crisis}
The capabilities of large language models (LLMs) have come at a substantial and rapidly growing resource cost. Training GPT-3 alone is estimated to have consumed approximately 1,287 MWh of electricity and emitted on the order of 552 tonnes of CO\textsubscript{2}-equivalent \cite{patterson2021carbon}. At the aggregate level, the trajectory is steeper still: the International Energy Agency projects that data-centre electricity consumption will roughly double from 485 TWh in 2025 to around 950 TWh by 2030, with consumption from AI-focused facilities tripling over the same period \cite{iea2026energy}. The scale of capital commitment is comparable, with the combined data-centre capital expenditure of the five largest technology companies exceeding USD 400 billion in 2025 \cite{iea2026energy}. Tellingly, the same analysis identifies a shortage of high-bandwidth memory---integral to AI chip production---that emerged in 2025 and is expected to persist through at least the end of 2027 \cite{iea2026energy}, a strong signal that memory, rather than logic, has become the binding constraint both economically and technically.

\begin{figure}[htbp]
\centering
\includegraphics[width=0.82\linewidth]{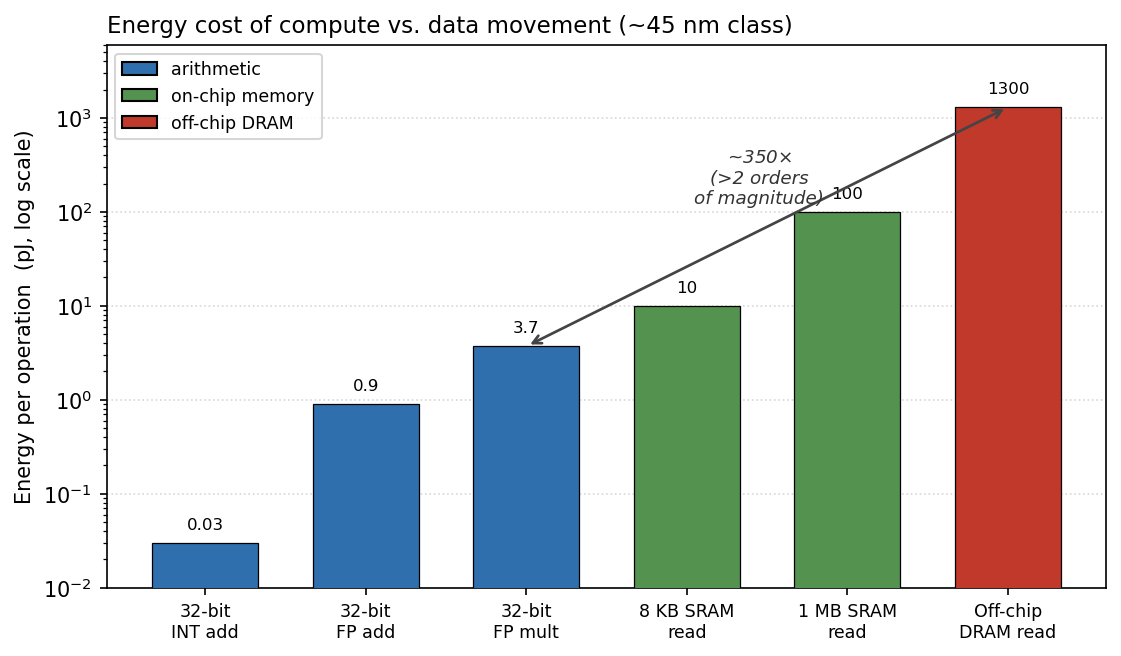}
\caption{Energy cost of arithmetic versus data movement in a modern process (\textasciitilde45 nm class). An off-chip DRAM access costs on the order of $10^{2}$--$10^{3}$\texttimes more energy than an on-chip arithmetic operation or SRAM access; values are representative order-of-magnitude figures after Horowitz \cite{horowitz2014computing}, with a single 64-bit DRAM access on the order of 1,300--2,600 pJ.}
\label{fig:energy}
\end{figure}

The deeper origin of this constraint is architectural. In modern process technologies, the energy required to move data dominates the energy required to compute on it: an off-chip DRAM access costs on the order of 1,300 to 2,600 picojoules per 64-bit access, against only a few picojoules for an on-chip floating-point operation---a gap of roughly three orders of magnitude that has not narrowed in proportion to logic scaling \cite{horowitz2014computing} (Figure \ref{fig:energy}). For the autoregressive decode phase of LLM inference, which repeatedly streams large weight matrices and a growing key--value cache from memory, this von Neumann bottleneck causes general-purpose accelerators to operate at only a fraction of their nominal peak throughput \cite{li2024llmhardware}. The efficiency challenge of LLM deployment is therefore, to first order, a problem of memory bandwidth and data movement rather than of raw arithmetic capability---a framing that motivates the hardware survey presented here.

\subsection{Why a Hardware Review Is Needed Now}
The hardware response to this challenge has accelerated markedly, with major vendors now committing to roughly annual accelerator releases. NVIDIA's progression from the Hopper to the Blackwell generation, and onward to the announced Rubin platform, has been paralleled by purpose-built silicon from the hyperscalers, including Google's seventh-generation Ironwood TPU, released in November 2025 and AWS Trainium3, the company's first 3 nm AI chip, which began shipping in December 2025, alongside specialized architectures such as the Cerebras WSE-3, the Groq LPU, and a growing cohort of start-ups---including d-Matrix and Etched---designing ASICs specifically for Transformer workloads \cite{kachris2025survey}. Despite this proliferation, GPUs are still estimated to account for roughly 80\% of AI-accelerator revenue---with NVIDIA alone projected to hold 70--75\% of the market through the decade \cite{bloomberg2026accel}---underscoring that diversity of supply has not yet translated into displacement of the incumbent.

This expanding design space---spanning general-purpose GPUs, domain-specific TPUs and ASICs, reconfigurable FPGAs, memory-centric processing-in-memory (PIM) and near-memory computing, and emerging neuromorphic and photonic devices---has outpaced the synthesizing literature. Existing surveys provide valuable foundations but tend to concentrate on inference alone, to treat a subset of hardware platforms, or to enumerate individual works without a unifying analytical lens \cite{kachris2025survey,li2024llmhardware,yuan2024llm}. This review addresses that gap through five contributions: (i) a taxonomy organizing the accelerator landscape by architecture, deployment context, and optimization target; (ii) a workload-grounded derivation of the hardware requirements that the Transformer imposes; (iii) a critical, per-paradigm survey from GPUs through PIM to neuromorphic and photonic approaches; (iv) a cross-paradigm comparison on absolute metrics---tokens per second and tokens per joule---anchored to a roofline analysis; and (v) a structured discussion of open challenges and research directions. In keeping with the present state of the evidence, we do not argue that GPUs will be displaced, but rather that the landscape is fragmenting into specialized niches and converging toward heterogeneous, memory-centric integration (Figure~\ref{fig:overview}).

\subsection{Scope, Methodology, and Organization}
The scope of this review encompasses GPUs, TPUs and custom ASICs, FPGAs, PIM and near-memory computing, and emerging neuromorphic and photonic devices, considered across three deployment contexts: cloud training, cloud inference, and edge inference. Non-Transformer architectures and pre-Transformer models are excluded except as background, on the grounds that Transformer-based models dominate deployed LLMs and define the workload of interest; where state-space models and mixture-of-experts designs materially alter the hardware profile, they are noted in context. The underlying literature was drawn from IEEE Xplore, the ACM Digital Library, arXiv, and ScienceDirect over the period 2018--2026, with priority given to work from 2022 onward, using search terms such as ``LLM accelerator,'' ``Transformer inference hardware,'' and ``processing-in-memory Transformer,'' and was supplemented by vendor technical documentation and MLPerf benchmark results where peer-reviewed data were unavailable. The remainder of the paper is organized as follows. Section 2 establishes the architectural and computational background and derives the resulting hardware requirements; Section 3 presents the taxonomy; Sections 4 through 10 survey GPU, ASIC, FPGA, PIM, neuromorphic, photonic, and edge accelerators in turn; Section 11 provides a unified comparative analysis; Section 12 examines open challenges and future directions; and Section 13 concludes.

\section{Background: LLM Architecture and Hardware Requirements}
\subsection{The Transformer Architecture}
Contemporary LLMs are predominantly decoder-only Transformers: a stack of identical layers, each pairing a multi-head attention (MHA) block with a position-wise feed-forward network (FFN), connected by residual paths and stabilized by layer or RMS normalization \cite{vaswani2017attention}. The defining operation is scaled dot-product attention over query, key, and value projections,
\begin{equation}
\text{Attention}(Q,K,V) = \text{softmax}\!\left(\frac{QK^{\top}}{\sqrt{d_k}}\right)V,
\end{equation}
in which the $QK^{\top}$ product produces an n \texttimes n score matrix for a sequence of length $n$, giving attention a computational and memory cost that scales as $O(n^2)$ in sequence length \cite{vaswani2017attention}. This quadratic term, negligible for short prompts, comes to dominate both latency and memory as context windows extend to tens or hundreds of thousands of tokens.

The way a model projects keys and values is, in practice, a memory-bandwidth decision rather than a modeling one. Under standard MHA each of the $h$ attention heads maintains its own key and value projections, so the cache of past keys and values that must be retained during autoregressive generation grows in proportion to the number of heads. Multi-query attention (MQA) shares a single key--value head across all query heads, reducing this cache by roughly a factor of $h$ at some cost to quality \cite{shazeer2019fast}. Grouped-query attention (GQA) interpolates between the two by organizing query heads into groups that each share one key--value head, lowering memory bandwidth and latency while preserving most of the quality of MHA \cite{ainslie2023gqa}, and it has consequently become the default in the Llama, Mistral, and Gemma model families \cite{ainslie2023gqa,dubey2024llama3,jiang2023mistral}. Multi-head latent attention (MLA) pushes the same logic further: introduced in DeepSeek-V2 and carried into the DeepSeek-V3 family, it compresses keys and values jointly into a low-rank latent vector and caches that latent in place of the full per-head tensors, cutting the KV cache by roughly an order of magnitude relative to MHA at comparable or better quality \cite{deepseek2024v2}. The progression from MHA through MQA and GQA to MLA is therefore best read as a hardware-driven trajectory, with each step purchasing decode bandwidth at the price of architectural simplicity. The remaining components---the FFN, which typically holds the majority of a dense model's parameters and floating-point operations; the embedding and output-projection layers, which scale with vocabulary size; and the elementwise normalization and residual operations---complete the layer and, as discussed below, exhibit markedly different hardware behavior from attention.

\subsection{The Computational Profile of Inference}
\begin{figure}[htbp]
\centering
\includegraphics[width=\linewidth]{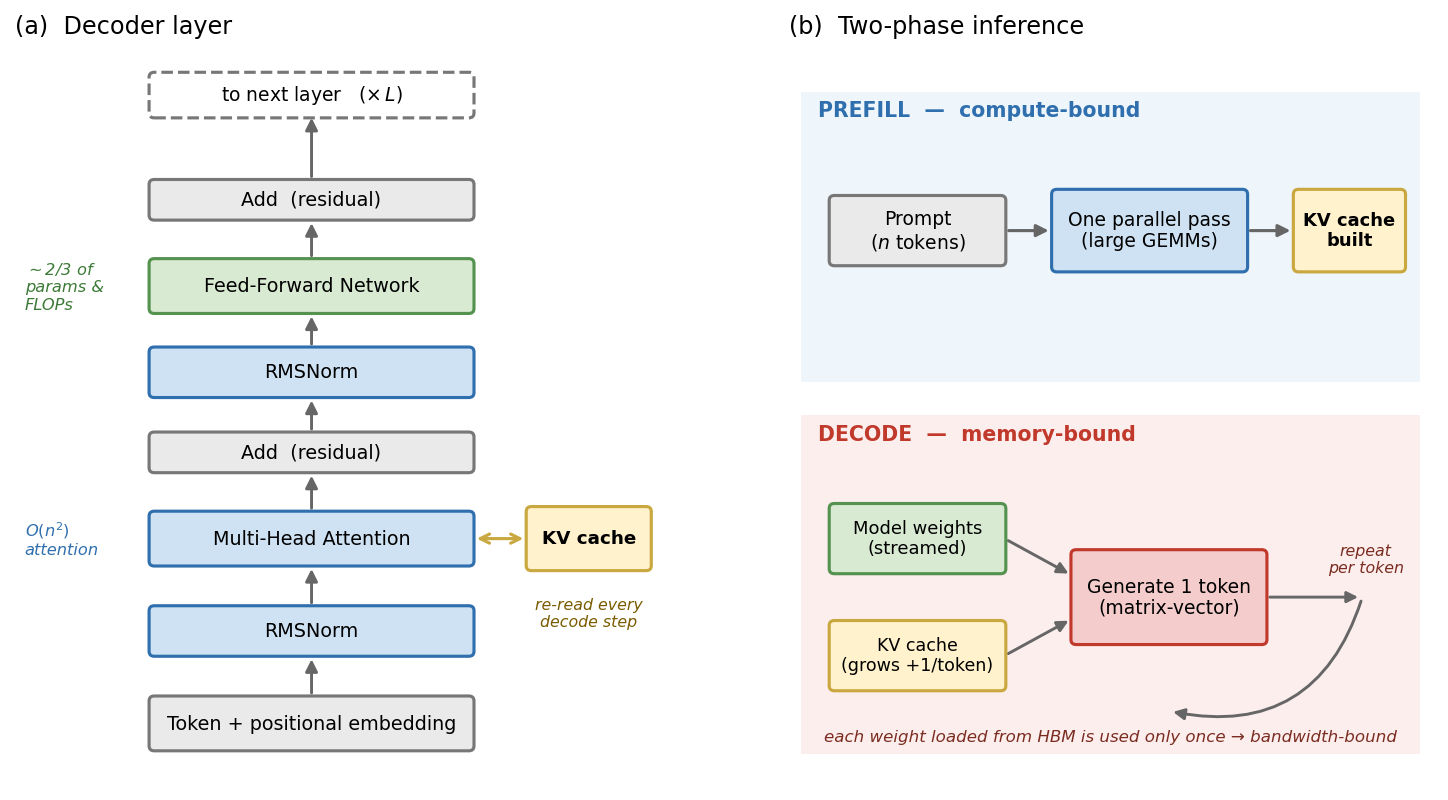}
\caption{Anatomy of LLM inference. (a) A decoder layer pairs multi-head attention---which maintains the key--value (KV) cache that is re-read at every decode step---with a feed-forward network holding the majority of parameters and FLOPs. (b) Inference runs in two phases: a compute-bound \emph{prefill} that processes the whole prompt in one parallel pass and builds the KV cache, and a memory-bound \emph{decode} that emits one token per step while streaming the weights and the growing KV cache from memory.}
\label{fig:transformer}
\end{figure}

LLM inference proceeds in two phases with fundamentally different bottlenecks (Figure \ref{fig:transformer}). During the \textit{prefill} phase the entire prompt is processed in parallel and the initial key--value cache is constructed; this phase is compute-bound, consisting of large matrix multiplications with substantial data reuse. During the \textit{decode} phase tokens are generated one at a time, each conditioned on the accumulated cache. Decode is memory-bound: the high-dimensional weight matrices loaded for a single generated token are used only once before the next step, leaving the arithmetic units idle \cite{pope2023efficiently}, and because every sequence in a batch carries its own key--value cache, attention during generation remains memory-bandwidth-bound regardless of batch size.

\begin{figure}[htbp]
\centering
\includegraphics[width=0.82\linewidth]{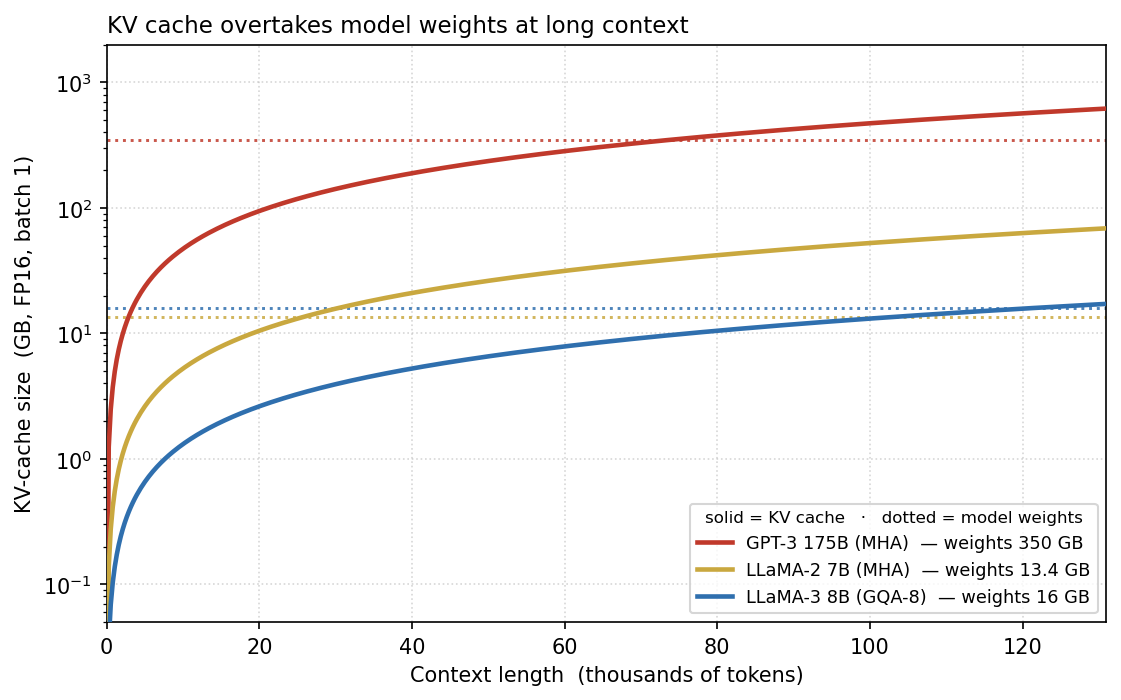}
\caption{Key--value cache size versus context length (batch 1, FP16), computed from the parameters in Table \ref{tab:models}. For multi-head-attention models the cache overtakes the model weights (dotted lines) at moderate context lengths, whereas grouped-query attention (LLaMA-3 8B) keeps it far smaller; the 128K range exceeds the native context window of some listed models and is shown only to illustrate scaling.}
\label{fig:kvcache}
\end{figure}

The key--value (KV) cache is therefore central. For a batch size of one in FP16, its size is $2 \cdot L \cdot h_{kv} \cdot d_{\text{head}} \cdot n_{\text{tokens}} \cdot 2$ bytes---linear in both context length and batch size---where $L$ is the layer count and $h_{kv}$ the number of key--value heads. As Table \ref{tab:models} and Figure \ref{fig:kvcache} show, at long context the cache rivals or exceeds the model weights themselves, and because it must be re-read at every decode step it, more than raw model size, sets the bandwidth ceiling on generation throughput. This single fact motivates a large fraction of the techniques surveyed later, from GQA and paged cache management \cite{kwon2023pagedattention} to KV-cache compression (Section 12).

These behaviors are made precise by the roofline model \cite{williams2009roofline}, which bounds achievable throughput by $\min\!\left(P_{\text{peak}},\, B \cdot I\right)$ is peak compute, $B$ is memory bandwidth, and $I$ is the arithmetic intensity of the workload:
\begin{equation}
I = \frac{\text{FLOPs}}{\text{bytes accessed}}, \qquad I_{\text{crit}} = \frac{P_{\text{peak}}}{B}.
\end{equation}
Operations whose intensity falls below the critical value $I_{\text{crit}}$ are bandwidth-bound; those above it are compute-bound. Decode attention has a low and nearly constant arithmetic intensity and thus sits far into the bandwidth-bound regime on essentially all current accelerators, whereas prefill and the FFN at large batch sizes are compute-bound \cite{pope2023efficiently}. This gap---between the floating-point throughput a device advertises and the fraction a bandwidth-bound decode can actually consume---is the analytical thread of this review, and is ultimately a manifestation of the data-movement energy asymmetry quantified by Horowitz \cite{horowitz2014computing} and the broader memory-wall trend \cite{gholami2024memorywall}. Much of the hardware examined in Sections 4--10 can be read as an attempt either to raise the bandwidth roof or to shift the workload's position relative to $I_{\text{crit}}$.

\begin{table}[htbp]
\centering
\caption{Computational and memory characteristics of representative LLMs (batch size 1, FP16).}
\label{tab:models}
\footnotesize
\setlength{\tabcolsep}{4pt}
\resizebox{\textwidth}{!}{
\begin{tabular}{lllrrrr}
\toprule
Model & Parameters & Architecture & Training FLOPs\textsuperscript{b} & Weights (FP16) & KV cache @4K\textsuperscript{a} & KV cache @128K\textsuperscript{a} \\
\midrule
GPT-3 & 175 B & Dense, MHA & 3.1\texttimes$10^{23}$ & 350 GB & 19.3 GB & 618 GB \\
GPT-4 (est.) & \textasciitilde1.8 T (reported) & Undisclosed (reportedly MoE) & \textasciitilde$10^{25}$ (est.) & --- & --- & --- \\
LLaMA-2 7B & 6.7 B & Dense, MHA & $\approx8.4$\texttimes$10^{22}$ & 13.4 GB & 2.1 GB & 68.7 GB \\
LLaMA-2 70B & 70 B & Dense, GQA (8 KV) & $\approx8.4$\texttimes$10^{23}$ & 140 GB & 1.3 GB & 42.9 GB \\
LLaMA-3 8B & 8 B & Dense, GQA (8 KV) & $\approx7.2$\texttimes$10^{23}$ & 16 GB & 537 MB & 17.2 GB \\
LLaMA-3 70B & 70 B & Dense, GQA (8 KV) & $\approx6.3$\texttimes$10^{24}$ & 140 GB & 1.3 GB & 42.9 GB \\
Gemma 2B & 2.5 B & Dense, MQA & --- & 5 GB & 75 MB & 2.4 GB \\
Mistral 7B & 7.3 B & Dense, GQA (8 KV) & --- & 14.6 GB & 537 MB & 17.2 GB \\
Mixtral 8\texttimes7B & 46.7 B total / 12.9 B active & Sparse MoE, GQA (8 KV) & --- & 93.4 GB & 537 MB & 17.2 GB \\
\bottomrule
\end{tabular}}
\\[4pt]
{\footnotesize \raggedright \textsuperscript{a} KV cache computed as $2\,L\,h_{kv}\,d_{\text{head}}\,n_{\text{tokens}}$\texttimes2 bytes; the 128K column illustrates scaling and exceeds the native context window of several listed models. \textsuperscript{b} LLaMA training FLOPs estimated via the $6ND$ heuristic with published token counts (2T for LLaMA-2, 15T for LLaMA-3); GPT-3 from \cite{brown2020language}; GPT-4 figures are public estimates, not confirmed; others undisclosed.\par}
\end{table}

\subsection{Hardware Requirements Derived from the Workload}
The profile above translates into a concrete and partly conflicting set of hardware requirements. The foremost is memory bandwidth: because the decode phase that dominates interactive serving is bandwidth-bound, sustained bandwidth---rather than peak floating-point throughput---governs achievable token rates. Closely related is the need for large, fast on-chip memory, since keeping weights and the KV cache in SRAM near the compute units avoids costly off-chip traffic, a principle taken to its logical extreme by the wafer-scale and SRAM-resident designs of Section 5. At the same time, efficient general matrix-multiply (GEMM) engines remain essential for the compute-bound prefill and FFN computations, as does native support for low-precision arithmetic---spanning BF16 and FP16 down through FP8 and INT8 to INT4---which reduces memory footprint, bandwidth pressure, and compute energy simultaneously. Sparsity support is increasingly required in two forms: regular, structured sparsity amenable to dense datapaths, and the irregular, data-dependent access patterns introduced by mixture-of-experts routing. Finally, because frontier models far exceed the memory of any single device, high-bandwidth interconnects for multi-chip and multi-node parallelism become a first-order design constraint rather than an afterthought. That these requirements pull in different directions---bandwidth against capacity, specialization against flexibility---is precisely why no single architecture dominates, and it motivates the taxonomy introduced in Section 3.

\subsection{Quantization and Sparsity as Hardware Levers}
Reduced-precision quantization is the most widely deployed software lever, and its hardware implications are substantial. Weight-only methods directly attack the dominant memory and bandwidth cost of decode: GPTQ compresses weights to three or four bits per parameter with minimal accuracy loss and can quantize a 175-billion-parameter model in a few GPU-hours using second-order information \cite{frantar2023gptq}, while AWQ instead protects the small fraction of salient weight channels identified by activation magnitude \cite{lin2024awq}. Activation quantization is considerably harder: beyond roughly 6.7 billion parameters, systematic activation outliers---often around one hundred times larger than typical values and confined to fixed channels---sharply degrade na\"ive INT8 quantization. LLM.int8() addresses this by isolating outlier channels in a higher-precision path \cite{dettmers2022llmint8}, whereas SmoothQuant migrates the quantization difficulty from activations to weights through a mathematically equivalent per-channel transformation, enabling 8-bit weight and 8-bit activation inference \cite{xiao2023smoothquant}; the FP8 formats now standard on recent training and inference hardware extend the same principle to floating point \cite{micikevicius2022fp8}. The collective hardware implication is that an accelerator targeting LLMs must natively support a range of numeric formats together with per-channel or per-group scaling; Table \ref{tab:quantmethods} summarizes the principal methods.

\begin{table}[htbp]
\centering
\caption{Principal quantization and sparsity techniques for LLMs and their hardware implications.}
\label{tab:quantmethods}
\footnotesize
\setlength{\tabcolsep}{5pt}
\renewcommand{\arraystretch}{1.15}
\begin{tabularx}{\textwidth}{>{\hsize=0.75\hsize\raggedright\arraybackslash}X >{\hsize=0.70\hsize\raggedright\arraybackslash}X >{\hsize=0.65\hsize\raggedright\arraybackslash}X >{\hsize=1.90\hsize\raggedright\arraybackslash}X}
\toprule
Technique & Class & Precision & Key idea / hardware implication \\
\midrule
GPTQ \cite{frantar2023gptq} & Weight-only PTQ & 3--4-bit weights & Second-order (Hessian) error compensation; quantizes a 175B model in a few GPU-hours \\
AWQ \cite{lin2024awq} & Weight-only PTQ & 3--4-bit weights & Protects the salient weight channels identified by activation magnitude \\
LLM.int8() \cite{dettmers2022llmint8} & Weight + activation & INT8 (+FP16 outliers) & Isolates systematic outlier channels in a higher-precision path \\
SmoothQuant \cite{xiao2023smoothquant} & Weight + activation & W8A8 & Migrates activation outliers into the weights via an equivalent per-channel scaling \\
FP8 (E4M3/E5M2) \cite{micikevicius2022fp8} & Weight + activation & 8-bit float & Hardware-native low precision on recent training and inference silicon \\
SpinQuant \cite{liu2024spinquant} & Weight (+ activation) & 4-bit & Learns rotation matrices that suppress outliers before quantization \\
2:4 sparsity \cite{mishra2021sparse} & Structured sparsity & 50\% (2 of 4) & Regular pattern with predictable acceleration on dense datapaths \\
Mixture-of-experts \cite{fedus2022switch,jiang2024mixtral} & Conditional compute & --- & Routes each token to a few experts; raises capacity but also memory footprint \\
\bottomrule
\end{tabularx}
\end{table}

Sparsity offers a second lever with a sharp hardware dichotomy. Structured patterns such as 2:4 sparsity map onto regular datapaths with predictable acceleration \cite{mishra2021sparse}, whereas the unstructured sparsity of attention and the dynamic, token-dependent routing of mixture-of-experts (MoE) models promise larger theoretical savings but impose irregular memory access that general-purpose hardware exploits poorly. MoE designs---from the Switch Transformer \cite{fedus2022switch} to Mixtral \cite{jiang2024mixtral}---decouple total model capacity from per-token compute, yet they raise aggregate memory pressure, since all experts must remain resident, and add routing and communication overhead. The unifying observation is that each of these software techniques alters the arithmetic intensity and memory-access pattern of the workload, so its benefit is realized only when the underlying hardware is co-designed to exploit it. This co-design lens is the one we carry through the remainder of the review.

\section{A Taxonomy of LLM Hardware Accelerators}
\begin{figure}[htbp]
\centering
\includegraphics[width=0.95\linewidth]{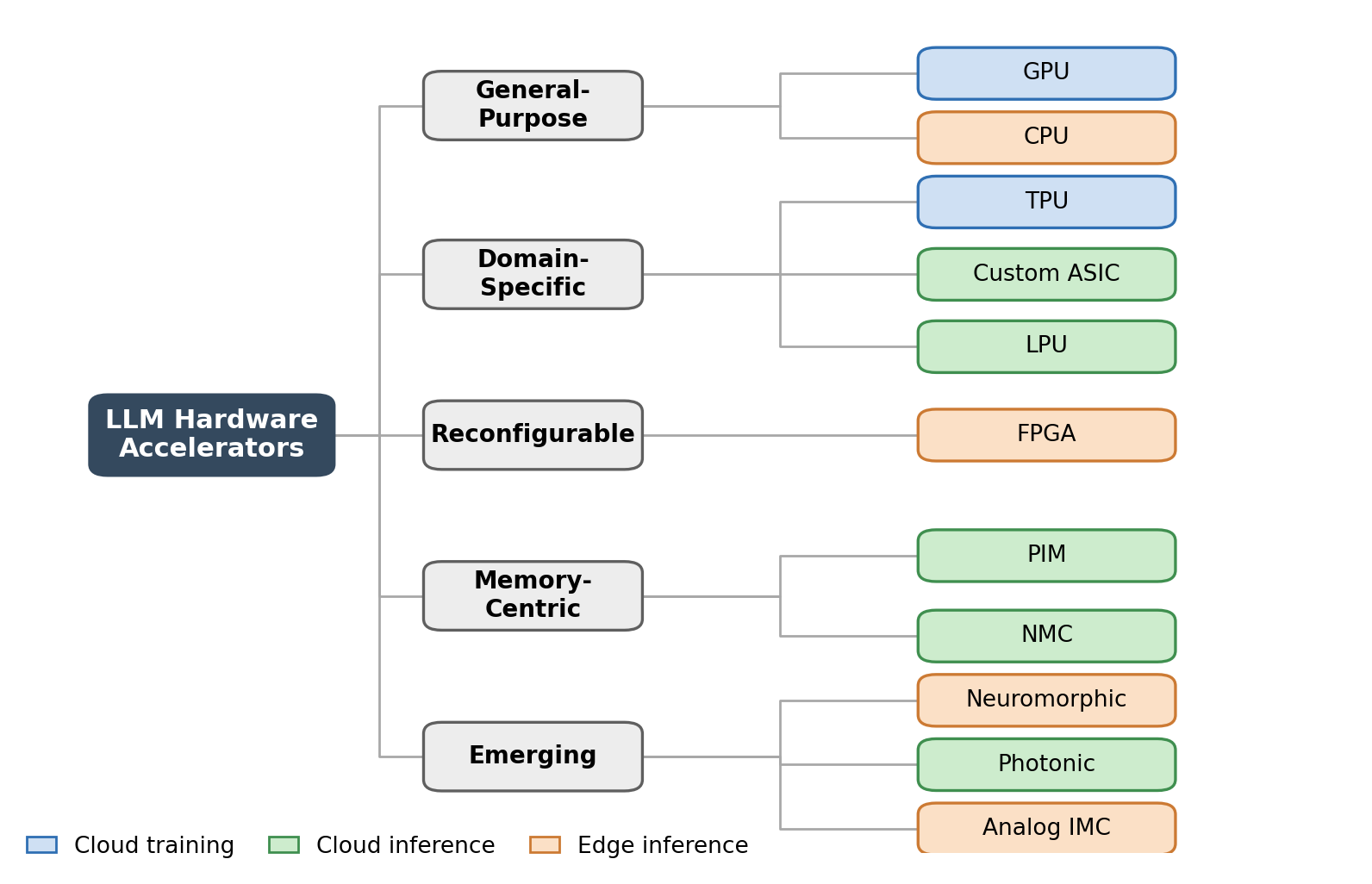}
\caption{Taxonomy of LLM hardware accelerators by architecture (tree structure) and, by colour, the deployment context in which each class is most commonly applied. Several classes span multiple contexts; colour denotes the most common one.}
\label{fig:taxonomy}
\end{figure}

The accelerator landscape is heterogeneous enough that meaningful comparison requires an explicit organizing framework. We classify LLM accelerators along three orthogonal axes---architecture, deployment context, and optimization target---and apply this scheme throughout the remainder of the review. Figure \ref{fig:taxonomy} summarizes the architectural axis and, by color, the deployment context in which each class is most commonly applied.

\subsection{By Architecture Type}
The architectural axis spans a spectrum that trades programmability against efficiency \cite{kachris2025survey,li2024llmhardware}. \textit{General-purpose} processors---GPUs and CPUs---offer the broadest workload coverage and the most mature software ecosystems. \textit{Domain-specific} accelerators---Tensor Processing Units (TPUs), custom application-specific integrated circuits (ASICs), and language processing units (LPUs)---sacrifice flexibility for higher efficiency on a narrower set of operations. \textit{Reconfigurable} devices, principally field-programmable gate arrays (FPGAs), occupy an intermediate position, allowing the datapath to be re-specialized after fabrication. \textit{Memory-centric} architectures, including processing-in-memory (PIM) and near-memory computing (NMC), attack the data-movement bottleneck directly by placing computation within or adjacent to the memory arrays. Finally, \textit{emerging} substrates---neuromorphic, photonic, and analog in-memory computing---depart from the conventional von Neumann model entirely and remain at comparatively early maturity. Each subsequent section of this review treats one of these classes in turn.

\subsection{By Deployment Context}
The same architectural class can serve very different roles, and the binding constraints differ sharply across them. \textit{Cloud training} prioritizes aggregate throughput and high-bandwidth interconnect across thousands of devices, operating at large batch sizes and mixed precision. \textit{Cloud inference} shifts the emphasis toward cost per token and latency under realistic serving loads. \textit{Edge inference} is instead bounded by power, area, and thermal envelopes, and is motivated by on-device privacy, offline operation, and the elimination of network round-trips. Because these contexts impose distinct requirements on an otherwise identical chip, deployment context is treated as a first-class dimension rather than a secondary attribute.

\subsection{By Optimization Target}
A given deployment is typically dominated by one of four metrics: throughput (tokens per second), latency (time-to-first-token and inter-token latency), energy efficiency (tokens per joule), and cost efficiency (tokens per dollar) \cite{li2024llmhardware,yuan2024llm}. These targets generally trade off against one another, so the notion of a single ``best'' accelerator is ill-posed; the appropriate choice is the device whose strengths align with the dominant metric of the intended use case. These three axes are not independent---a memory-centric device may be optimized for energy efficiency in an edge setting, for instance---and Section 11 situates concrete hardware within this combined space and distills the result into a deployment-oriented recommendation matrix (Table \ref{tab:matrix}).

\section{GPU-Based Acceleration}
\subsection{Why GPUs Became the Default}
GPUs remain the de facto platform for both LLM training and inference, for reasons that are as much ecosystem as architecture. Their single-instruction-multiple-thread (SIMT) execution model maps naturally onto the dense matrix multiplications at the core of the Transformer, and successive generations have added Tensor Cores---dedicated mixed-precision matrix-multiply units---together with high-bandwidth memory (HBM) to feed them \cite{kachris2025survey}. The decisive advantage, however, is the maturity of the CUDA software stack and its surrounding ecosystem of libraries, compilers, and frameworks, which lowers the barrier to deployment relative to any competing platform and creates substantial switching costs \cite{li2024llmhardware}. The result is an incumbency that, as the comparisons below show, persists even where competing hardware matches or exceeds raw specifications.

\subsection{NVIDIA Architecture Evolution}
Across four generations, NVIDIA's data-center GPUs trace the shifting bottleneck of LLM workloads. The Ampere A100 (2020) established the template with 80 GB of HBM2e at roughly 2.0 TB/s, third-generation Tensor Cores delivering 312 dense BF16 TFLOPS, and a 400 W power envelope \cite{nvidia2021a100}. The Hopper H100 (2022) added a Transformer Engine with FP8 support and reached approximately 990 dense BF16 TFLOPS on HBM3 at 3.35 TB/s \cite{nvidia2023h100}. The H200 left compute unchanged but raised memory to 141 GB of HBM3e at 4.8 TB/s---an upgrade aimed squarely at the memory-bound decode phase rather than at peak throughput \cite{nvidia2023h100}. The Blackwell generation (B100/B200) introduced fifth-generation Tensor Cores with native FP4 and FP6, 192 GB of HBM3e at 8 TB/s and a fifth-generation NVLink providing 1.8 TB/s per GPU at a TDP of up to 1,000 W for discrete form factors and configurable up to 1,200 W for liquid-cooled GB200 configurations; at the rack scale, NVIDIA reports that a GB200 NVL72 system generates roughly thirty times more inference tokens than an equivalent Hopper configuration \cite{nvidia2024blackwell}. Table \ref{tab:gpu} collects representative specifications.

\begin{table}[htbp]
\centering
\caption{Representative data-center GPUs for LLM workloads (BF16 figures are dense; structured sparsity roughly doubles them).}
\label{tab:gpu}
\scriptsize
\setlength{\tabcolsep}{3pt}
\resizebox{\textwidth}{!}{
\begin{tabular}{lllllrrll}
\toprule
GPU & Vendor & Process / die & Memory & Bandwidth & TDP & BF16 dense & Tensor core & Price\textsuperscript{a} \\
\midrule
A100 80GB SXM & NVIDIA & 7 nm / 826 mm\textsuperscript{2} & 80 GB HBM2e & 2.0 TB/s & 400 W & 312 TFLOPS & 3rd gen & \$10--15k \\
H100 SXM5 & NVIDIA & 4N / 814 mm\textsuperscript{2} & 80 GB HBM3 & 3.35 TB/s & 700 W & \textasciitilde990 TFLOPS & 4th gen & \$25--30k \\
H200 SXM & NVIDIA & 4N / 814 mm\textsuperscript{2} & 141 GB HBM3e & 4.8 TB/s & 700 W & \textasciitilde990 TFLOPS & 4th gen & \$30--35k \\
B100 & NVIDIA & 4NP / dual-die & 192 GB HBM3e & 8 TB/s & 700 W & \textasciitilde1,800 TFLOPS & 5th gen & \textasciitilde\$30k \\
B200 & NVIDIA & 4NP / dual-die (208 B tr.) & 192 GB HBM3e & 8 TB/s & 1,000 W & \textasciitilde2,250 TFLOPS & 5th gen & \$30--40k \\
MI300X & AMD & 5/6 nm chiplet (153 B tr.) & 192 GB HBM3 & 5.3 TB/s & 750 W & 1,307 TFLOPS & CDNA 3 Matrix & \$15--20k \\
Gaudi 3 & Intel & 5 nm / dual-die & 128 GB HBM2e & 3.7 TB/s & 900 W & \textasciitilde1,835 TFLOPS & MME & \textasciitilde\$15k \\
\bottomrule
\end{tabular}}
\\[4pt]
{\footnotesize \raggedright \textsuperscript{a} Prices are indicative street/list estimates and fluctuate substantially; sources: \cite{nvidia2021a100,nvidia2023h100,nvidia2024blackwell,amd2023mi300x,intel2024gaudi3}.\par}
\end{table}

A recurring source of confusion in such comparisons is the dense-versus-sparse distinction: vendor headline figures typically assume structured sparsity, so that Blackwell's widely quoted FP4 throughput of roughly 18--20 PFLOPS corresponds to about half that in dense operation. We report dense values throughout for comparability.

\subsection{GPU Bottlenecks for LLM Inference}
Despite their dominance, GPUs are poorly matched to the decode phase. As established in Section 2, decode is memory-bandwidth-bound, so the Tensor Cores that define a GPU's peak throughput sit largely idle while the weights and the KV cache are streamed from HBM \cite{pope2023efficiently}. The inefficiency is concrete: even an optimized attention kernel such as FlashAttention-2 achieves only about 35\% utilization on the H100, against 80--90\% for a well-tuned matrix multiply \cite{shah2024flashattention3}. Single-user and small-batch serving make this worse, since there is little weight reuse over which to amortize memory traffic, and the KV cache imposes additional management overhead \cite{kwon2023pagedattention}. At the multi-device scale that frontier models require, communication across tensor- and pipeline-parallel partitions adds further overhead that can dominate at high degrees of parallelism \cite{shoeybi2019megatron}.

\subsection{Software Optimizations}
A substantial body of work narrows this gap in software, and its effectiveness is itself evidence that the bottleneck is structural rather than computational. The FlashAttention family reorganizes attention into an IO-aware, tiled, fused kernel that avoids materializing the full attention matrix in HBM: the original reaches only 25--40\% of peak, while FlashAttention-2 improves work partitioning to reach 50--73\% of peak on the A100 \cite{dao2022flashattention,dao2024flashattention2}, and FlashAttention-3 further exploits asynchronous execution and low precision on Hopper \cite{shah2024flashattention3}. At the serving layer, continuous batching dynamically packs requests to raise device utilization \cite{yu2022orca}, and PagedAttention manages the KV cache in non-contiguous pages to reduce fragmentation and support longer contexts \cite{kwon2023pagedattention}. Speculative decoding attacks the sequential nature of generation directly: a small draft model proposes multiple tokens that the target model verifies in a single forward pass, yielding speedups proportional to the acceptance rate while preserving the output distribution \cite{leviathan2023speculative}. Finally, tensor, pipeline, and sequence parallelism distribute both computation and memory across devices, making trillion-parameter training tractable \cite{shoeybi2019megatron}.

A complementary serving-level response follows directly from the two-phase structure of Section 2.2: because prefill is compute-bound and decode bandwidth-bound, co-locating both phases on the same GPU forces each to interfere with the other's resource profile. Disaggregated serving instead schedules prefill and decode on separate device pools, transferring the KV cache between them over high-bandwidth interconnect; Splitwise \cite{patel2024splitwise} and DistServe \cite{zhong2024distserve} report substantial gains in throughput per GPU and in goodput under latency constraints from this split, and the pattern has since entered production inference stacks. Disaggregation is notable here because it operationalizes, at cluster scale, exactly the workload partition that motivates the heterogeneous GPU-plus-PIM designs of Section 7---further evidence that the prefill/decode asymmetry, not raw throughput, is the organizing fact of LLM serving.

\subsection{AMD and Other Competitors}
Competing GPUs increasingly match or exceed NVIDIA on raw specifications, particularly memory. AMD's MI300X pairs 192 GB of HBM3 with 5.3 TB/s of bandwidth and a marketed 1,307 dense BF16 TFLOPS, sufficient to hold a 70-billion-parameter model on a single device \cite{amd2023mi300x}. Intel's Gaudi 3 takes a scale-out-oriented approach with 128 GB of HBM2e at 3.7 TB/s and roughly 1.8 PFLOPS of BF16/FP8 matrix compute \cite{intel2024gaudi3}. Yet specification parity has not translated into displacement, and the reason is software: in independent benchmarking the MI300X reached only about 620 BF16 TFLOPS against its 1,307 marketed figure---slower in practice than the H100 despite a higher nominal peak---owing to the relative immaturity of AMD's ROCm communication and library stack \cite{semianalysis2024mi300x}. This gap between nominal and realized performance, more than any single architectural feature, explains why the CUDA ecosystem remains the field's center of gravity, and it motivates the specialized, non-GPU approaches examined in the sections that follow.

\section{Custom ASIC Accelerators}
\subsection{Google TPU Family}
The Tensor Processing Unit, the longest-running custom AI ASIC, is built around a systolic-array matrix-multiply unit first detailed for the inference-only v1 \cite{jouppi2017tpu} and since scaled across generations. TPU v4 introduced an optically reconfigurable interconnect and delivered 275 dense BF16 TFLOPS per chip in 4,096-chip pods \cite{jouppi2023tpuv4}. The v5 generation then split into a cost-optimized v5e and a performance-oriented v5p, the latter providing 459 BF16 TFLOPS and 95 GB of HBM2e at 2.76 TB/s, scaling to 8,960-chip pods, and used to train Gemini models. The seventh-generation Ironwood (v7), generally available since late 2025, marks an inflection toward inference: it is the first TPU with native FP8, delivering 4,614 FP8 (2,307 BF16) TFLOPS and 192 GB of HBM3e at 7.37 TB/s per chip, with 9,216-chip superpods reaching 42.5 FP8 ExaFLOPS at roughly 600 W per chip \cite{google2025ironwood}. The trajectory is unambiguous---each generation prioritizes memory capacity, bandwidth, and inference efficiency---and the announced eighth generation reportedly splits into separate training and inference chips, an explicit acknowledgment that the two workloads have divergent hardware requirements.

\subsection{Groq LPU}
The Groq Language Processing Unit embodies the opposite philosophy. Built on a Tensor Streaming Processor architecture with fully deterministic, compiler-scheduled execution \cite{abts2020tsp}, it eliminates runtime arbitration and stores all data in on-chip SRAM rather than HBM: each chip contains roughly 230 MB of SRAM with about 80 TB/s of on-die bandwidth---an order of magnitude beyond HBM3e---but cannot hold even a small model alone. The architectural consequence is stark and was understated in early accounts: serving Llama-2/3 70B requires on the order of 576 LPUs spread across roughly nine racks, which together deliver about 300 tokens per second per user---roughly ten times an H100---at an energy cost near 1--3 joules per token. The design thus trades capacity and chip count for latency determinism, a tradeoff compelling enough that in December 2025 NVIDIA entered a non-exclusive licensing agreement for Groq's inference technology---reported at roughly \$20 billion, the largest transaction in NVIDIA's history---under which Groq's founder and core engineering team joined NVIDIA \cite{groq2025nvidia}.

\subsection{Cerebras Wafer-Scale Engine}
Cerebras pursues the SRAM-resident philosophy at the largest scale physics permits: the whole wafer. The WSE-3 is a single wafer-scale device of 46,225 mm\textsuperscript{2} containing over four trillion transistors, 900,000 cores, and 44 GB of on-chip SRAM at 21 PB/s of aggregate bandwidth, with a published peak of 125 PFLOPS---a figure that refers to sparse FP16, Cerebras having disclosed no dense equivalent \cite{cerebras2024wse3}. By keeping an entire model on one piece of silicon, the design eliminates chip-to-chip communication for single-model training, and an external memory subsystem of up to 1.2 PB allows the CS-3 system to hold models of up to 24 trillion parameters in a single logical space. Its commercial relevance has grown sharply: in January 2026 OpenAI announced a multi-year agreement, valued at over \$10 billion, under which Cerebras will supply 750 MW of wafer-scale inference capacity through 2028, aimed in particular at low-latency serving of long-output reasoning workloads \cite{openai2026cerebras}---a substantial validation of the wafer-scale bet for high-throughput serving.

\subsection{AWS Inferentia and Trainium}
Amazon's strategy is vertical integration for cost efficiency. Inferentia2, built on the NeuronCore-v2 architecture, targets inference with configurable FP8 support and roughly 50\% better performance-per-watt than comparable instances, while the Trainium line addresses both training and inference. Trainium2 delivers 667 dense BF16 (1,299 FP8) TFLOPS with 96 GB of HBM3e at 2.9 TB/s in an approximately 500 W envelope, and underpins Project Rainier, a cluster of roughly 500,000 chips used to train and serve Anthropic's Claude models. The newer Trainium3, AWS's first 3 nm chip, doubles compute to 2.52 FP8 PFLOPS and raises memory to 144 GB of HBM3e at 4.9 TB/s. The consistent claim across this line is economic rather than peak-performance leadership: AWS reports on the order of 30--40\% better price-performance than comparable GPU instances and, internally, roughly 54\% lower cost per token for GPT-class serving \cite{aws2024neuron}. Table \ref{tab:asic} collects the custom ASICs surveyed in this section.

\begin{table}[htbp]
\centering
\caption{Representative custom AI ASICs (dense figures unless noted).}
\label{tab:asic}
\scriptsize
\setlength{\tabcolsep}{3pt}
\resizebox{\textwidth}{!}{
\begin{tabular}{llllllll}
\toprule
Chip & Vendor & Memory & Capacity & Bandwidth & Peak compute & TDP & Primary target \\
\midrule
TPU v5p & Google & HBM2e & 95 GB & 2.76 TB/s & 459 BF16 TFLOPS & --- & Training \\
TPU v7 (Ironwood) & Google & HBM3e & 192 GB & 7.37 TB/s & 2,307 BF16 / 4,614 FP8 TFLOPS & \textasciitilde600 W & Training + inference \\
LPU v1 & Groq & On-chip SRAM & 230 MB\textsuperscript{a} & 80 TB/s (on-chip) & 188 FP16 TFLOPS / 750 INT8 TOPS & --- & Low-latency inference \\
WSE-3 (CS-3) & Cerebras & On-chip SRAM & 44 GB & 21 PB/s (on-chip) & 125 PFLOPS (sparse FP16)\textsuperscript{b} & --- & Training \\
Trainium2 & AWS & HBM3e & 96 GB & 2.9 TB/s & 667 BF16 / 1,299 FP8 TFLOPS & \textasciitilde500 W & Training + inference \\
Trainium3 & AWS & HBM3e & 144 GB & 4.9 TB/s & 2,520 FP8 TFLOPS & --- & Training + inference \\
AI200 & Qualcomm & LPDDR & 768 GB / card & LPDDR\textsuperscript{c} & not disclosed & 160 kW (rack) & Inference (capacity) \\
\bottomrule
\end{tabular}}
\\[4pt]
{\footnotesize \raggedright \textsuperscript{a} Per chip; a 70B model is served across hundreds of LPUs. \textsuperscript{b} Sparse FP16; no dense figure published. \textsuperscript{c} LPDDR favours capacity over bandwidth; the AI250 successor targets $>$10\texttimes effective bandwidth via near-memory computing. TDP ``---'' where vendors do not publish a per-chip figure.\par}
\end{table}

\subsection{Tenstorrent}
Tenstorrent stakes out an open-hardware position. Its processors are built from Tensix cores---each comprising five ``baby'' RISC-V microprocessors alongside a matrix unit, SIMD unit, and local SRAM---and span Grayskull, Wormhole, and the current Blackhole, which offers 745 FP8 (372 FP16) TFLOPS, 32 GB of GDDR6, and an Ethernet-based interconnect across 140 Tensix cores \cite{tenstorrent2024}. The distinguishing bet is a fully open-source stack, principally the low-level TT-Metalium SDK. Yet the practical obstacle mirrors the GPU discussion of Section 4: as of early 2026, much of the verified model support and tooling targets the older Wormhole cards, FlashAttention-3 has no port, and there is no production equivalent to vLLM's continuous-batching serving API, leaving software maturity as the binding constraint.

\subsection{Qualcomm}
Qualcomm extends its mobile NPU lineage into the data center with an explicitly memory-centric, inference-only design. The AI200 (2026) provides 768 GB of LPDDR per card---roughly an order of magnitude more capacity than an H100, at lower cost and bandwidth than HBM---betting that inference is capacity- rather than bandwidth-limited, while the AI250 (2027) adopts a near-memory computing architecture claiming over ten times the effective memory bandwidth \cite{qualcomm2025ai200}. This wager that LPDDR capacity can substitute for HBM bandwidth in autoregressive serving, and the AI250's near-memory direction in particular, anticipates the processing-in-memory approaches examined in Section 7.

\subsection{SambaNova, Graphcore, and Others}
Several reconfigurable-dataflow architectures complete the landscape. SambaNova's SN40L pairs a reconfigurable dataflow unit with a three-tier memory hierarchy of on-chip SRAM, HBM, and direct-attached DDR, and a composition-of-experts approach; it reportedly serves a 70B model on just 16 chips, in sharp contrast to Groq's hundreds \cite{prabhakar2024sambanova}. Graphcore's Intelligence Processing Unit, by contrast, used fine-grained MIMD parallelism with large distributed on-chip SRAM and bulk-synchronous-parallel execution, but struggled commercially and was acquired in 2024 \cite{jia2019graphcore}. Taken together, the custom-ASIC field separates into two camps---training-throughput designs (TPU, Trainium, WSE) and inference-latency or inference-cost designs (Groq, Qualcomm, Inferentia)---but every one of them is, at root, an attempt to attack the memory bottleneck, whether by maximizing on-chip SRAM, expanding HBM capacity and bandwidth, or moving computation toward memory. That common thread motivates the memory-centric architectures of Section 7 and the cross-paradigm comparison of Section 11; first, however, Section 6 examines the middle path between GPU generality and ASIC commitment---a datapath that can be respecialized after fabrication.

\section{FPGA-Based Acceleration}
\subsection{Why FPGAs for LLMs}
Field-programmable gate arrays occupy the middle ground between general-purpose GPUs and fixed-function ASICs, offering a datapath that can be re-specialized after fabrication. This reconfigurability is valuable precisely because LLM architectures continue to churn---grouped-query attention, mixture-of-experts, and state-space variants each alter the ideal datapath---and an FPGA can be retargeted without a new silicon respin. The fabric's abundant DSP blocks map naturally onto custom quantized matrix multiplication, its spatial dataflow avoids the instruction-fetch overhead of a programmable processor, and its heterogeneous on-chip memory (block and ultra RAM) can be arranged to keep activations close to compute. For small to medium deployment scales, these properties translate into lower power and a lower cost of entry than a GPU cluster, and the approach has datacenter precedent: Microsoft's Catapult program integrated FPGAs across its fleet to accelerate large-scale services \cite{putnam2014catapult}.

\subsection{Key Research and Deployments}
Microsoft demonstrated FPGA acceleration at production scale with Project Brainwave, which served deep neural networks in real time for Bing's intelligent search and Azure by exploiting distributed model parallelism and low-latency hardware microservices \cite{chung2018brainwave}. For LLMs specifically, recent work maps transformers onto HBM-equipped FPGA cards. DFX, a multi-FPGA appliance using four Xilinx Alveo U280 cards with model parallelism and optimized dataflow across both the prefill and decode phases, reaches roughly 120 tokens per second on GPT-2 1.5B \cite{hong2022dfx}. The more complete FlightLLM provides an end-to-end mapping flow built on a configurable sparse DSP chain and an always-on-chip decode scheme with mixed-precision support; implemented on the Alveo U280, it achieves 6.0\texttimes higher energy efficiency and 1.8\texttimes better cost efficiency than an NVIDIA V100S on LLaMA-2 7B, and beats an A100's throughput by 1.2\texttimes on the newer Versal VHK158, at a batch size of one \cite{zeng2024flightllm}. On the toolchain side, streaming-dataflow frameworks such as FINN \cite{umuroglu2017finn} and vendor high-level-synthesis flows lower models onto the fabric---though, as noted below, this remains the principal friction. Table \ref{tab:fpga} summarizes representative FPGA LLM accelerators.

\begin{table}[htbp]
\centering
\caption{Representative FPGA-based accelerators for LLM and transformer inference.}
\label{tab:fpga}
\footnotesize
\setlength{\tabcolsep}{5pt}
\renewcommand{\arraystretch}{1.15}
\begin{tabularx}{\textwidth}{>{\hsize=0.70\hsize\raggedright\arraybackslash}X >{\hsize=1.00\hsize\raggedright\arraybackslash}X >{\hsize=0.80\hsize\raggedright\arraybackslash}X >{\hsize=1.50\hsize\raggedright\arraybackslash}X}
\toprule
Work & Platform & Model / task & Reported result \\
\midrule
Brainwave \cite{chung2018brainwave} & Intel Stratix 10 & Production DNN serving & Real-time, low-latency serving at datacenter scale \\
DFX \cite{hong2022dfx} & 4\texttimes Alveo U280 & GPT-2 1.5B & \textasciitilde120 tokens/s; multi-FPGA model parallelism over prefill and decode \\
FlightLLM \cite{zeng2024flightllm} & Alveo U280 / Versal VHK158 & LLaMA-2 7B & 6.0\texttimes energy- and 1.8\texttimes cost-efficiency vs.\ V100S; 1.2\texttimes A100 throughput on VHK158 \\
FTRANS \cite{li2020ftrans} & Xilinx VCU118 & Transformer LM & Block-circulant weight compression with little accuracy loss \\
FINN \cite{umuroglu2017finn} & Framework (HLS) & Quantized networks & Streaming-dataflow generation of quantized inference \\
\bottomrule
\end{tabularx}
\end{table}

\subsection{Quantization-Aware FPGA Design}
FPGAs are most competitive where aggressive compression and custom numeric formats align with the fabric. INT4 and INT8 datapaths map efficiently onto DSP and lookup-table resources, and block floating point is a particularly natural fit: pioneered in Microsoft's MSFP for Brainwave, it shares a single exponent across a block of values, and achieves accuracy comparable to BF16 and INT8 at roughly 3\texttimes and 4\texttimes lower cost respectively, with negligible accuracy loss and no change to model topology, across CNNs, RNNs, and transformers \cite{rouhani2020msfp}. Structured-weight compression has also been applied directly to FPGA transformer inference; FTRANS, for example, uses block-circulant matrix representations to compress transformer language models at the algorithm level with little accuracy degradation \cite{li2020ftrans}. The broader point is that FPGA flexibility allows the numeric format and the sparsity pattern to be co-matched to the available silicon---exactly the principle that FlightLLM's configurable sparse DSP chain operationalizes---making FPGAs a strong vehicle for hardware-aware, co-designed inference.

\subsection{Limitations of FPGAs for LLMs}
These advantages are bounded by the same memory wall that constrains every other platform, and on FPGAs it binds harder. Even an HBM-equipped Alveo U280 provides only about 460 GB/s of HBM bandwidth alongside 38 GB/s of DDR---well below the multi-terabyte-per-second bandwidth of a contemporary GPU (Table \ref{tab:ceilings}). FPGAs therefore cannot stream weights and the KV cache quickly enough to compete on large-batch cloud serving, nor approach an H100 or B200 on training throughput. The second limitation is programmability: high-level synthesis, register-transfer-level design, and long place-and-route cycles raise the engineering barrier substantially, echoing the software-maturity theme that recurs throughout this review. The realistic niche for FPGAs is therefore edge and embedded inference, and latency-critical, low-batch, single-user serving where determinism and energy efficiency outweigh peak throughput---precisely the regime that DFX and FlightLLM target. Within that niche FPGAs are a genuine complement to GPUs and ASICs rather than a replacement, and their tight coupling of custom precision to a reconfigurable fabric anticipates the memory-centric designs examined next.

\section{Processing-in-Memory and Near-Memory Computing}
\subsection{The Memory Wall}
The architectures of Sections 4--6 differ in datapath but share a constraint: data must travel between separate memory and compute units. This von Neumann separation is the origin of the memory wall, and its cost is fundamentally one of energy and bandwidth. As established in Section 2, a DRAM access consumes roughly two orders of magnitude more energy than an on-chip operation \cite{horowitz2014computing}, and memory bandwidth has scaled far more slowly than compute across hardware generations \cite{gholami2024memorywall}. Because the decode phase that dominates LLM serving is bandwidth-bound, and because off-package bandwidth is physically capped by pin count, board wiring, and thermal limits, simply widening the memory interface offers diminishing returns. Processing-in-memory (PIM) and near-memory computing instead attack the problem at its source by moving computation toward the data.

\subsection{PIM Fundamentals}
Three categories are usefully distinguished. \textit{In-DRAM PIM} places compute units inside the DRAM device, adjacent to the banks, exploiting the very high internal bank bandwidth that is otherwise throttled at the chip's external interface. \textit{Processing-near-memory} (PNM) instead places compute units near but outside the DRAM arrays---on a logic die or in the memory controller or buffer---trading some bandwidth for easier integration. \textit{Processing-using-memory} exploits the analog behavior of the memory array itself to perform computation in place. The central tradeoff is density: near-bank processing units fabricated in the DRAM process impose area overhead that reduces memory density, which is especially detrimental for the large memory footprints of LLMs, whereas PNM preserves density at lower internal bandwidth.

\subsection{Industry PIM Chips}
Both major DRAM vendors have built PIM silicon. Samsung's HBM-PIM (Aquabolt-XL) integrates a 16-lane FP16 SIMD compute unit next to pairs of DRAM banks in HBM2 layers, exposing roughly 4.92 TB/s of in-DRAM compute bandwidth against 1.23 TB/s at the external interface---a 4\texttimes internal advantage. It ships as a drop-in JEDEC-compatible device with a software stack that runs unmodified TensorFlow and PyTorch code \cite{kim2021aquabolt}, and a 2023 extension targeted generative AI workloads including GPT, T5, and Llama-2. SK Hynix's Accelerator-in-Memory (AiM) is a GDDR6-based design placing an array of FP16 multipliers, an adder tree, an accumulator, and lookup-table-based activation logic alongside each bank, with a small global SRAM buffer, delivering 1 TFLOPS and 1 TB/s per chip \cite{lee2022aim}; its AiMX scale-out prototype achieves about 330 tokens per second on OPT-6.7B at batch size one. UPMEM, by contrast, ships general-purpose PIM DIMMs with simple processor cores beside DDR memory \cite{devaux2019upmem}. Reported LLM results are encouraging within their scope---an HBM2-PIM system on an Alveo U280 board reaches roughly 348 tokens per second on GPT-1.3B, about 1.6\texttimes an A100, for the memory-bound GEMV portions of inference---but the constrained per-chip compute capability of DRAM-PIM remains its key limitation.

\subsection{Academic PIM Architectures for LLMs}
\begin{figure}[htbp]
\centering
\includegraphics[width=0.9\linewidth]{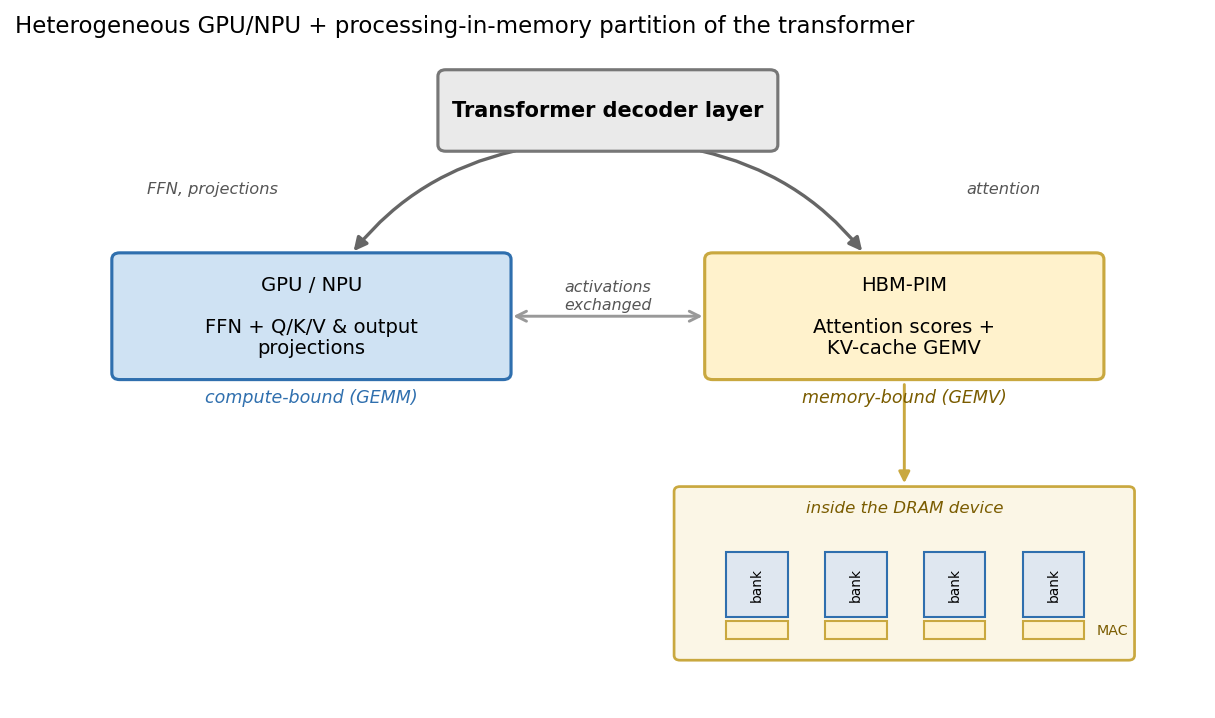}
\caption{Heterogeneous partition of the transformer across a GPU/NPU and processing-in-memory, the pattern adopted by designs such as AttAcc \cite{park2024attacc} and NeuPIMs \cite{heo2024neupims}: compute-bound feed-forward and projection GEMMs run on the GPU or NPU, while the memory-bound attention and KV-cache GEMV run on HBM-PIM, where compute units sit beside the DRAM banks.}
\label{fig:pim}
\end{figure}

Research designs sharpen the workload partitioning that this limitation implies. The motivating observation is specific to attention: because each request produces its own KV matrices, attention has a low operations-per-byte ratio regardless of batch size, and the aggregate KV cache can exceed the size of the model weights themselves, making it an ideal PIM target while the compute-bound feed-forward layers are better left on a conventional processor. TransPIM proposed a monolithic HBM-PIM with co-designed dataflows for the full transformer, though it was oriented toward encoder blocks and single-request inference \cite{zhou2022transpim}. AttAcc instead places only the attention layer on HBM-based PIM within a heterogeneous GPU-plus-PIM system, improving the performance and energy efficiency of a 175B model by up to 2.81\texttimes and 2.67\texttimes respectively over a conventional system of the same memory capacity \cite{park2024attacc}. NeuPIMs extends this to batched decoder inference by enabling an NPU and HBM-PIM to execute simultaneously rather than contending for the same device \cite{heo2024neupims}. The convergent architecture across these works---an NPU or GPU for compute-bound GEMM paired with PIM for memory-bound attention---reflects a broader consensus on how PIM most plausibly enters the LLM pipeline (Figure \ref{fig:pim}).

\subsection{Near-Memory Compute and CXL}
Near-memory designs place logic on the base die of a stacked memory or on a memory-attached device. Tesla's Dojo offers an instructive case: its D1 tile coupled large distributed on-die SRAM with a mesh interconnect in a dataflow training architecture \cite{talpes2023dojo}, but the effort illustrates the risk of bespoke silicon---Tesla wound down the Dojo program in 2025, with Musk describing the second generation as an ``evolutionary dead end'' and redirecting effort to its AI5/AI6 chips, though reports in early 2026 suggested a partial, board-level revival. Compute Express Link (CXL) has emerged as the more durable near-memory vector, providing memory expansion and pooling for LLM serving; PNM designs accordingly place compute units in memory controllers or CXL devices, and an LPDDR-based CXL-PNM platform has been proposed for cost-efficient inference of transformer LLMs \cite{park2024cxlpnm}. Because CXL-attached capacity directly addresses the growing KV cache, it connects the memory-centric story here to the GPU memory specifications of Table \ref{tab:gpu} and the bandwidth hierarchy of Figure \ref{fig:memhier}.

\subsection{Challenges and Open Problems}
Several obstacles temper near-term deployment. DRAM process and timing constraints limit the frequency and complexity of in-memory logic; near-bank compute reduces memory density at precisely the moment LLMs demand more capacity; and maintaining memory consistency across PIM and host introduces system-level complexity. As with every alternative platform in this review, programmability is the recurring bottleneck---commercial PIM exposes limited instruction sets and bespoke software stacks---and performance is acutely sensitive to data layout, motivating dedicated layout-optimization work. Taken together, the evidence supports a measured conclusion: PIM and near-memory computing are most credible not as GPU replacements but as heterogeneous complements that absorb the memory-bound attention and GEMV work, and the parallel industry moves toward near-memory designs---including Qualcomm's AI250 noted in Section 5.6---make this among the most promising near-term responses to the memory wall, a thread resumed in Section 12. The two sections that follow consider more radical departures, in which memory and compute are co-located by construction rather than by integration.

\section{Neuromorphic Chips for LLM Workloads}
\subsection{Neuromorphic Computing Primer}
Neuromorphic processors depart from the von Neumann model entirely. Rather than continuous floating-point activations, spiking neural networks (SNNs) communicate through discrete, asynchronous spikes, so computation is event-driven and occurs only when a neuron fires. Memory and compute are co-located---each neuron's state sits beside the synaptic weights that update it---which eliminates the data-movement bottleneck that motivates the rest of this review. The efficiency case rests on sparsity: because most neurons are silent at any instant, energy is expended only on active spikes, a regime well matched to the sparse activation patterns that also appear in large transformers.

\subsection{Key Neuromorphic Hardware}
The landmark devices span a decade of progress. IBM's TrueNorth (2014) \cite{merolla2014truenorth} integrated one million neurons and 256 million synapses while drawing roughly 65 mW, establishing the low-power, event-driven template. Intel's Loihi 2 advanced programmability: a single chip supports up to one million neurons and 120 million synapses, with programmable neuron models, graded spikes, and asynchronous event-driven communication \cite{davies2018loihi}, and at system scale Intel's Hala Point, deployed at Sandia in 2024, packs 1.15 billion neurons across 1,152 Loihi 2 processors. IBM's more recent NorthPole takes a near-memory rather than strictly spiking approach, co-locating memory and compute to reach up to 5x the energy efficiency of an H100 on image-recognition tasks \cite{modha2023northpole}. At the edge, BrainChip's Akida is a fully digital SNN system-on-chip fabricated in 28 nm with on-chip learning at milliwatt power for wearables and embedded use \cite{brainchip2021akida}. China's efforts illustrate both the field's scale ambitions and its architectural breadth: the rack-scale Darwin Monkey packs over two billion neurons and one hundred billion synapses across 960 custom chips while drawing only about 2,000 W, and the earlier hybrid Tianjic chip unified artificial and spiking neural networks on a single architecture \cite{pei2019tianjic}.

\subsection{Neuromorphic Approaches for Transformers}
Two routes connect this hardware to transformer workloads. The first is ANN-to-SNN conversion \cite{rueckauer2017conversion}, which replaces a pre-trained network's ReLU activations with spiking neurons; this preserves accuracy but typically requires many simulation timesteps to approximate the original activations, at a corresponding cost in latency. The second is direct training with surrogate gradients. Spiking attention has been the key architectural advance: Spikformer's Spiking Self-Attention drops the softmax and operates directly on spike-form queries, keys, and values, avoiding multiplications and yielding low operation counts and energy \cite{zhou2023spikformer}. For language specifically, SpikeGPT is the first fully spiking generative language model, built on the linear-attention RWKV architecture at 46M and 216M parameters and reporting roughly 5\texttimes lower energy on neuromorphic hardware that exploits its event-driven activations \cite{zhu2023spikegpt}. These results establish feasibility, but the spiking-transformer literature for NLP remains small relative to its vision counterpart.

\subsection{Research and Commercial Momentum}
Interest is accelerating on several fronts, though it is worth separating genuine progress from marketing. On the research side, a clear line runs from Spikformer through SpikeGPT to more recent scaling efforts. On the systems side, the billion-neuron Hala Point and Darwin Monkey machines demonstrate that neuromorphic hardware now scales to substantial sizes. Commercially, IBM has moved NorthPole toward production, BrainChip's Akida was evaluated by NASA for potential spaceflight use, its IP licensed by Frontgrade Gaisler for space-grade SoCs, and the company has raised funding for an ``Akida GenAI'' effort aiming to run roughly 1.2-billion-parameter LLMs entirely on-device. Device-level research into analog and memristive synapses continues to push the underlying technology, but these remain early-stage relative to digital accelerators.

\subsection{Honest Assessment: Neuromorphic for LLMs}
A candid appraisal is warranted, because the gap between promise and present capability is wide. No current neuromorphic chip can run a frontier-scale LLM: the largest demonstrated spiking language models are two to three orders of magnitude smaller than today's deployed models, and the on-device targets being announced sit in the low single-digit billions of parameters. The realistic near-term role is therefore edge inference of small or distilled models, where the energy and latency advantages of event-driven computation matter most and absolute model quality matters least. The longer-term potential is genuinely high---sparse transformers and SNN-transformer hybrids align naturally with neuromorphic execution---but it is contingent on advances in training spiking models at scale and in software maturity, the same constraint seen throughout this review. For mainstream LLM inference today, neuromorphic computing is best characterized as a promising research direction rather than a competitive platform, a sober conclusion that the comparative analysis of Section 11 reinforces.

\section{Photonic and Optical Computing}
\subsection{Photonic Computing Fundamentals}
Optical computing proposes to perform the linear algebra at the heart of neural networks using light rather than electrons. Photonic integrated circuits offer higher bandwidth and lower loss than electrical circuits, making them well suited to the massively-connected linear layers of neural networks, and several optical architectures have been proposed to exploit this---Mach--Zehnder interferometer (MZI) meshes, microring weight banks, and photonic crossbar arrays. The MZI mesh is the mainstream structure for optical matrix-vector multiplication, programmed by applying voltages to thermo-optic phase shifters, while microring resonators implement weights in a broadcast-and-weight scheme \cite{tait2017neuromorphic}. Beyond these guided-wave approaches, free-space diffractive architectures form a distinct branch, in which a stack of passive diffractive layers performs inference as light propagates through them \cite{lin2018alloptical}; the broader landscape of photonic AI hardware is surveyed by Shastri et al.\ \cite{shastri2021photonics}, with photonic-neural-network and optics-informed deep-learning fundamentals reviewed by Giamougiannis et al.\ \cite{giamougiannis2024photonic}. Wavelength-division multiplexing adds a further axis of parallelism by carrying independent signals on distinct wavelengths through the same waveguide.

\subsection{Why Photonics for LLMs}
The appeal for LLMs follows from the workload's structure. Because matrix multiplication dominates both prefill and the feed-forward layers, and because an optical mesh performs that multiplication as light propagates through it, the linear operation itself executes at the speed of light and at very low energy once the weights are encoded. Combined with the throughput that wavelength multiplexing offers, this is in principle attractive for the GEMM-heavy transformer, where a single optical pass could replace many electronic multiply-accumulate cycles. Whether this advantage transfers to transformers specifically has been studied directly: Anderson et al.\ analyzed ``optical transformers,'' modeling the accuracy and energy scaling of executing transformer models on optical hardware and finding that the energy benefit is realized chiefly at large model and batch sizes, where the optical multiply outweighs the fixed cost of electro-optical conversion \cite{anderson2024optical}.

\subsection{Current Research and Commercial Activity}
The foundational demonstration came in 2017, when Shen et al.\ realized an on-chip silicon MZI neural network capable of recognizing spoken vowels, with an external subsystem configuring the matrix elements for the optical vector-matrix multiplication \cite{shen2017deeplearning}. Two results in 2021 then established that integrated photonics can reach trillions of multiply-accumulate operations per second: Feldmann et al.\ built a photonic tensor core---an optical analogue of an ASIC---performing parallelized in-memory computing with phase-change-material weight arrays driven by an optical frequency comb \cite{feldmann2021parallel}, and Xu et al.\ reported an 11-TOPS photonic convolutional accelerator using a frequency-comb source for wavelength parallelism \cite{xu2021photonic}. Commercial activity has since grown substantially, but its center of gravity is instructive. Lightmatter, valued at \$4.4 billion and having raised \$850 million as of October 2025, develops both a photonic AI accelerator (Envise) and a photonic interconnect platform (Passage)---yet its recent products are overwhelmingly on the interconnect side: the Passage M1000, a 3D photonic interposer announced in 2025, provides 114 Tbps of total optical bandwidth and connectivity to thousands of GPUs in a single package \cite{lightmatter2025passage}. The same pattern holds across the field. Celestial AI's Photonic Fabric targets ultra-high-bandwidth, low-latency memory transactions to decouple memory from compute \cite{celestial2025photonic}, and Ayar Labs focuses on optical I/O to replace electrical SerDes, with backing from NVIDIA, AMD, and Intel. The companies pursuing optical computation have, in practice, found their nearest market in optical communication. Table \ref{tab:emerging} collects representative neuromorphic and photonic hardware discussed in this and the preceding section.

\begin{table}[htbp]
\centering
\caption{Representative neuromorphic and photonic hardware for AI workloads.}
\label{tab:emerging}
\footnotesize
\setlength{\tabcolsep}{5pt}
\renewcommand{\arraystretch}{1.15}
\begin{tabularx}{\textwidth}{>{\hsize=0.80\hsize\raggedright\arraybackslash}X >{\hsize=1.00\hsize\raggedright\arraybackslash}X >{\hsize=1.50\hsize\raggedright\arraybackslash}X >{\hsize=0.70\hsize\raggedright\arraybackslash}X}
\toprule
Device / system & Class & Key scale or specification & Status \\
\midrule
TrueNorth \cite{merolla2014truenorth} & Neuromorphic (digital SNN) & 1M neurons, 256M synapses, ~65 mW & Research (2014) \\
Loihi 2 \cite{davies2018loihi} & Neuromorphic & 1M neurons, 120M synapses per chip; on-chip learning & Research platform \\
Hala Point \cite{davies2018loihi} & Neuromorphic system & 1.15B neurons across 1,152 Loihi 2 chips & Deployed (Sandia, 2024) \\
NorthPole \cite{modha2023northpole} & Near-memory inference & Up to 5x H100 energy efficiency (vision) & Toward production \\
Akida \cite{brainchip2021akida} & Edge neuromorphic & 28 nm SoC, milliwatt-class, on-chip learning & Commercial (IP) \\
MZI mesh ONN \cite{shen2017deeplearning} & Photonic (coherent) & On-chip optical vowel recognition & Research (2017) \\
Photonic tensor core \cite{feldmann2021parallel} & Photonic + phase-change & Trillions of MAC/s with optical comb source & Research (2021) \\
Lightmatter Passage \cite{lightmatter2025passage} & Photonic interconnect & 114 Tbps total optical I/O & Early product \\
\bottomrule
\end{tabularx}
\end{table}

\subsection{Challenges}
The reasons for this are physical, not merely commercial. The most fundamental is precision: analog optical matrix-vector multiplication is intrinsically limited in numerical precision by accumulated noise in electro-optical processing, and reaching 16-bit precision has required digital-analog hybrid architectures that reintroduce digitization \cite{zhou2025digital}. Nonlinearity is a second obstacle, since optical activation functions are difficult to realize and signals are typically converted back to the electronic domain for the nonlinear step, incurring conversion overhead. That conversion is itself the dominant cost: the modulators, digital-to-analog and analog-to-digital converters, and photodetectors that bridge the electronic and optical domains consume most of the energy and area and bound the achievable throughput, so the cheap optical multiply is surrounded by expensive interfaces. Microring-based schemes additionally suffer resonance drift with temperature, requiring careful calibration and adding power overhead. Above all, the central limitation of integrated photonic neural networks is scalability---encoding the vast number of network parameters onto a photonic chip---which has motivated tiled matrix-multiplication techniques that virtually enlarge a small physical circuit. And with no practical optical memory, the weights and KV cache of an LLM must reside in electronic memory in any case---so even a perfect optical multiplier would inherit the electronic memory wall intact.

\subsection{The Realistic Near-Term Role: Co-Packaged Optics}
Taken together, these constraints place optical computation for frontier LLMs firmly in the research stage. The credible near-term contribution of photonics is therefore not to replace the GPU's datapath but to relieve the data-movement bottlenecks that limit large-model systems---precisely the memory wall and inter-chip bandwidth limits identified in Section 7. Co-packaged optics, photonic interposers, and optical memory fabrics address exactly this, and the volume of capital and engineering now flowing into photonic interconnect rather than photonic compute reflects an industry consensus on where light is most useful today. Optical computation may yet mature into a role in inference, but for the present the bankable photonic win for LLM infrastructure is moving data, a judgment the comparative analysis of Section 11 adopts.

\section{Edge and Mobile LLM Deployment}
\subsection{Edge Constraints}
Edge deployment inverts every assumption of the datacenter. A phone or laptop offers a power budget of roughly one to ten watts rather than hundreds, no active cooling, and memory measured in single-digit gigabytes shared with the operating system and other applications. Crucially, mobile memory bandwidth---tens to perhaps a hundred gigabytes per second of LPDDR---is one to two orders of magnitude below the multi-terabyte-per-second HBM of Section 4, which makes the memory-bound decode phase even more acute on device than in the cloud (Table \ref{tab:ceilings} quantifies the gap). What motivates on-device inference despite these constraints is the combination of privacy, since data never leaves the device; latency, since there is no network round-trip; offline availability; and cost, since no server capacity is consumed.

\subsection{Mobile NPUs and SoCs}
Every major mobile SoC now integrates a dedicated neural processing unit. Apple's A18 Pro carries a 16-core Neural Engine rated at 35 trillion operations per second \cite{apple2024a18}, though notably Apple's Neural Engine has remained at roughly 35 TOPS since 2023, with the company emphasizing memory bandwidth and tight ecosystem integration rather than raw throughput. Qualcomm has pushed harder on peak numbers: the Snapdragon 8 Elite's Hexagon NPU delivers about a 45\% increase in AI performance over its predecessor \cite{qualcomm2024snapdragon}, and its PC-class Snapdragon X Elite NPU reaches 45 TOPS. MediaTek's Dimensity 9400 uses the NPU 890, which the company claims offers 80\% faster LLM prompt performance along with on-device LoRA training and video generation \cite{mediatek2024dimensity}. Nearly all of these NPUs sit alongside Arm CPU and GPU IP, which supplies the baseline compute platform across the mobile market. The TOPS figures themselves are vendor-stated and measured under differing conditions, and should be read as rough capability classes rather than directly comparable benchmarks.

\subsection{On-Device Models}
The change that made mobile LLMs practical was less about silicon than about models: small models became good enough. Apple's on-device foundation model is a roughly 3-billion-parameter model designed to run on an iPhone 15 Pro, exposed to developers through the Foundation Models framework in iOS 26, using aggressive low-bit palettization rather than the uniform 2-bit scheme sometimes reported \cite{gunter2024apple}, and it performs favorably against the larger Qwen-2.5-3B and Gemma-3-4B while remaining competitive with Qwen-3-4B. Meta released Llama 3.2 1B and 3B explicitly for edge use \cite{meta2024llama32}, quantizing them with both quantization-aware training plus LoRA adaptors and a post-training method, with inference supported in PyTorch's ExecuTorch framework on Qualcomm and MediaTek SoCs---the post-training method being SpinQuant, which learns rotation matrices to suppress the outliers that otherwise wreck low-bit accuracy \cite{liu2024spinquant,pytorch2024executorch}. Microsoft's Phi-3-mini (3.8B) \cite{abdin2024phi3} and Google's Gemma 2B \cite{gemma2024} occupy the same niche. By 2026, flagship phones from Samsung, Google, and Motorola supported on-device inference for models up to roughly 4B parameters in 4-bit quantization \cite{xu2024ondevice}. Architectural research has reinforced this trajectory: MobileLLM showed that below one billion parameters, deep-thin architectures outperform wide-shallow ones, and compute-optimal inference strategies let a Llama 3.2 1B with tree search outperform the 8B model \cite{liu2024mobilellm}. What does not yet exist is a mixture-of-experts design practical within a mobile envelope of roughly 10 W and 8 GB---a gap revisited in Section 12.

\subsection{Quantization and Edge Frameworks}
Aggressive quantization is the enabling technique, and the levers introduced in Section 2---GPTQ and AWQ among them---are precisely what bring multi-billion-parameter models within a mobile memory budget, with newer rotation-based methods such as SpinQuant further closing the low-bit accuracy gap \cite{liu2024spinquant}; 4-bit weights are the practical workhorse. A maturing software stack supports this: ExecuTorch \cite{pytorch2024executorch} and llama.cpp \cite{gerganov2023llamacpp} for portable inference, Apple's Core ML and Foundation Models framework, MLC-LLM \cite{mlc2023mlcllm}, and ONNX Runtime, alongside Apple's MLX framework, which is optimized specifically for Apple Silicon \cite{apple2024mlx}. A practical caveat is that much on-device LLM inference still runs on the CPU or GPU rather than the NPU, since NPU toolchains for the dynamic shapes and attention patterns of LLMs remain less mature than for the convolutional workloads they were designed around---an instance of the programmability gap seen throughout this review.

\subsection{The Memory Bottleneck on Edge}
The memory wall of Section 7 reappears here in miniature and, if anything, more sharply. Model weights must fit in RAM shared with everything else the device is doing; the KV cache grows with context length against a far smaller budget; and LPDDR bandwidth, not arithmetic throughput, sets the token-generation rate. The mitigations mirror the cloud---low-bit weight and KV-cache quantization---alongside edge-specific approaches such as offloading mixture-of-experts weights to flash storage and fetching each expert only when its router activates it. The constraint is fundamental rather than incidental: on a memory-bound workload, the device with less bandwidth generates tokens more slowly regardless of its peak TOPS.

\subsection{Use Cases and Outlook}
The realistic division of labor is hybrid. Small on-device models handle latency-sensitive and privacy-sensitive tasks---summarization, autocomplete, rewriting, and increasingly on-device agents---while harder queries escalate to the cloud, the pattern Apple embodies by pairing its on-device model with a server-based model in Private Cloud Compute. The trajectory is genuine and already shipping at the scale of a few billion parameters, but the appropriate conclusion is bounded: edge silicon has made small-model inference ubiquitous, while frontier-scale models remain a datacenter workload for the foreseeable future. Section 11 places these mobile parts in quantitative context against their cloud counterparts.

\section{Comparative Analysis}
\subsection{Dimensions of Comparison}
Comparing accelerators for LLMs is complicated by the fact that no single number captures fitness for the task. Six dimensions recur throughout the preceding sections: peak compute throughput, memory capacity, memory bandwidth, energy efficiency, software maturity, and scalability, with cost shadowing all of them. The weight each dimension carries is workload-dependent. Training and the prefill phase are compute- and bandwidth-intensive at high arithmetic intensity; the decode phase is bandwidth-bound at low intensity; latency-critical serving rewards low per-token latency, whereas throughput-critical serving rewards aggregate tokens per second per watt. A part that excels in one regime can be mediocre in another, which is why the sections above declined to rank platforms outright. This section accordingly compares the design space rather than crowning a winner.

\subsection{The Roofline View}
\begin{figure}[htbp]
\centering
\includegraphics[width=0.9\linewidth]{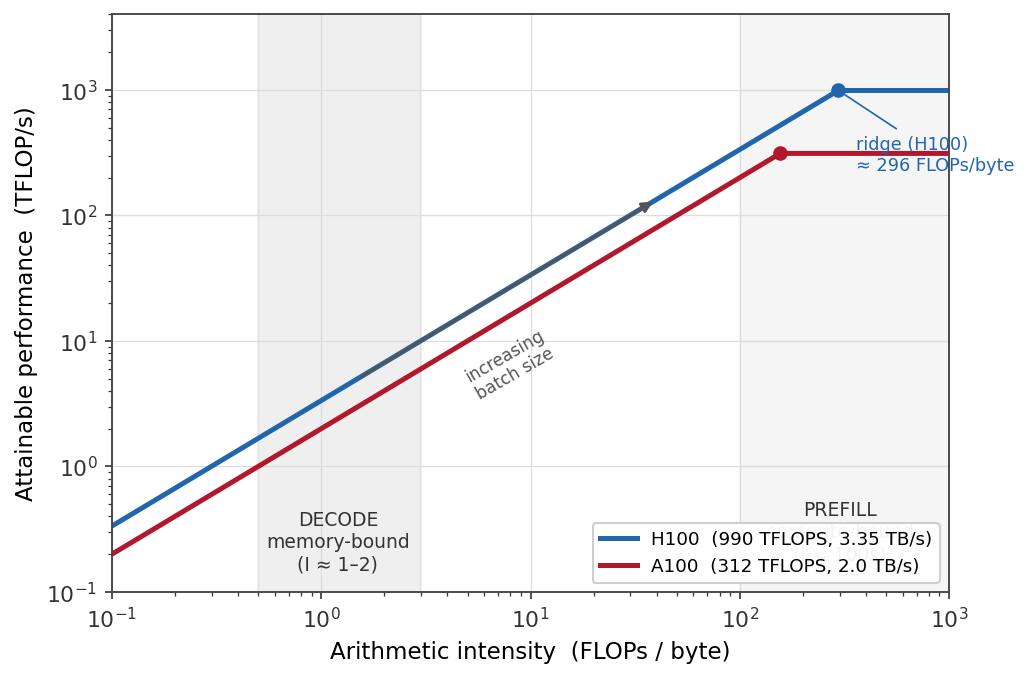}
\caption{Roofline model contrasting the A100 and H100. The memory-bound decode phase (arithmetic intensity $I \approx 1$--2 FLOPs/byte) sits far below peak throughput, whereas compute-bound prefill approaches the ceiling; increasing batch size shifts decode rightward.}
\label{fig:roofline}
\end{figure}

The roofline model introduced in Section 2 makes the central asymmetry of LLM inference visible. Plotting attainable performance against arithmetic intensity, every accelerator has a sloped memory-bound region set by its bandwidth and a flat compute-bound ceiling set by its peak throughput, meeting at a ridge point at intensity equal to peak FLOPS divided by bandwidth \cite{williams2009roofline}. Prefill, with its large matrix multiplications, sits at high intensity near or beyond the ridge and can approach peak compute; single-token decode is a sequence of matrix-vector products at an intensity of roughly one to two FLOPs per byte, placing it far down the memory-bound slope where attainable performance is a small fraction of peak \cite{yuan2024llm}. Batching shifts decode rightward by reusing each loaded weight across more tokens, which is precisely why throughput-oriented serving batches aggressively. The generational comparison in Figure \ref{fig:roofline} also shows that moving from the A100 to the H100 raised both the ceiling and the slope, but raised the ridge intensity as well, so a larger share of the decode regime remains bandwidth-limited.

\subsection{From Bandwidth to Tokens: First-Order Decode Ceilings}
The roofline view can be made concrete in the unit that serving economics actually prices: tokens per second. In the bandwidth-bound decode regime every generated token requires streaming, at minimum, the model weights once, so a first-order upper bound on single-stream throughput is simply $T_{\max} \approx B / S_{\text{model}}$, where $B$ is memory bandwidth and $S_{\text{model}}$ is the weight footprint at the serving precision. Table \ref{tab:ceilings} applies this bound to representative platforms from Sections 4--10 for a 70-billion-parameter model at FP16 and INT4 and an 8-billion-parameter model at INT4. Three observations follow. First, no quantity of additional FLOPS changes these numbers: even the highest-bandwidth HBM device cannot exceed a few tens of tokens per second per user for a 70B model at FP16, which is precisely why the SRAM-resident designs of Section 5 exist. Second, 4-bit weight quantization raises every ceiling fourfold---a software change worth more than a hardware generation, since the A100-to-B200 transition raised bandwidth by about the same factor over two generations---which is why low-bit weights are the single most consequential serving-side optimization (Section 2.4). Third, the same arithmetic reproduces the edge experience of Section 10: an 8B model at INT4 against roughly 77 GB/s of LPDDR5X is bounded near 19 tokens per second, squarely in the range observed on current flagship handsets, making explicit that on-device generation speed is set by LPDDR bandwidth rather than NPU TOPS. Measured rates fall below these bounds by the cost of KV-cache reads---which grow with context length---attention computation, and scheduling overheads; the bound's value is diagnostic, separating what better kernels can recover from what only more bandwidth can buy.

\begin{table}[htbp]
\centering
\caption{First-order upper bounds on single-stream (batch-1) decode throughput, computed as memory bandwidth divided by weight footprint\textsuperscript{a}. Values in tokens per second.}
\label{tab:ceilings}
\footnotesize
\setlength{\tabcolsep}{5pt}
\resizebox{\textwidth}{!}{
\begin{tabular}{lrrrr}
\toprule
Accelerator & Bandwidth & 70B @ FP16 (140 GB) & 70B @ INT4 (35 GB) & 8B @ INT4 (4 GB) \\
\midrule
NVIDIA H100 SXM5 & 3.35 TB/s & 24\textsuperscript{b} & 96 & 838 \\
NVIDIA H200 & 4.8 TB/s & 34 & 137 & 1,200 \\
NVIDIA B200 & 8.0 TB/s & 57 & 229 & 2,000 \\
AMD MI300X & 5.3 TB/s & 38 & 151 & 1,325 \\
Google TPU v7 (Ironwood) & 7.37 TB/s & 53 & 211 & 1,843 \\
AWS Trainium2 & 2.9 TB/s & 21\textsuperscript{b} & 83 & 725 \\
AMD Alveo U280 (HBM) & 0.46 TB/s & 3.3\textsuperscript{b} & 13\textsuperscript{b} & 115 \\
Flagship phone (LPDDR5X) & 0.077 TB/s & n/a & n/a & 19 \\
\bottomrule
\end{tabular}}
\\[4pt]
{\footnotesize \raggedright \textsuperscript{a} Idealized bound: batch size 1, weights-only memory traffic; ignores KV-cache reads (which grow with context length), attention computation, kernel and scheduling overheads, and communication. \textsuperscript{b} Weights exceed single-device memory; the bound then applies to the aggregate bandwidth of a sharded deployment. ``n/a'': the model does not fit on the device and sharding is not applicable.\par}
\end{table}

\subsection{The Compute--Bandwidth Design Space}
\begin{figure}[htbp]
\centering
\includegraphics[width=0.9\linewidth]{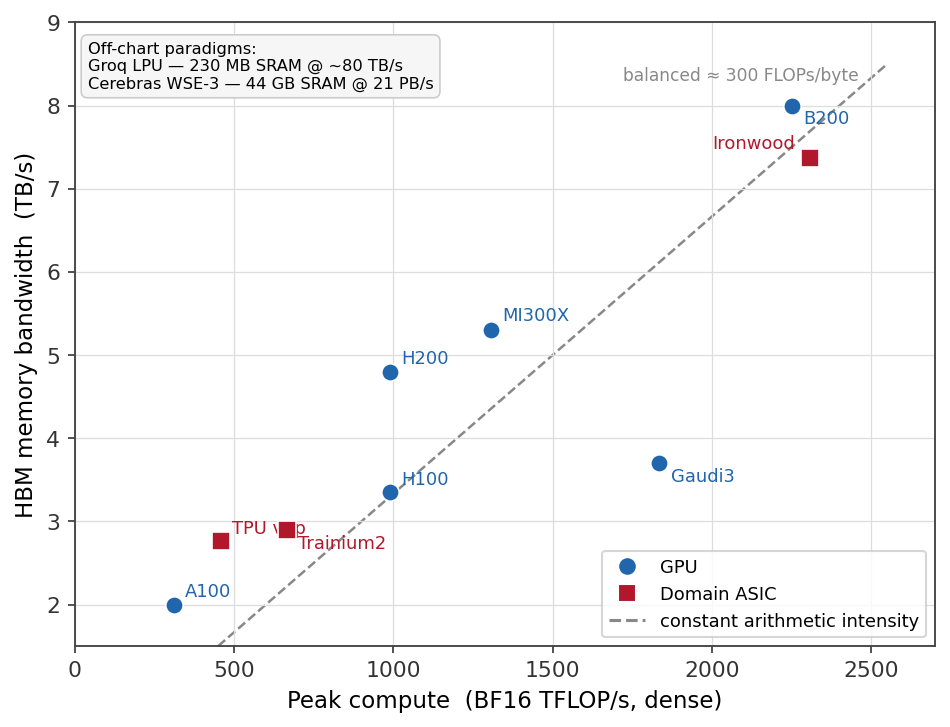}
\caption{Compute--bandwidth design space of representative datacenter accelerators (nameplate dense BF16). The dashed line marks a constant arithmetic intensity of \textasciitilde300 FLOPs/byte. SRAM-resident designs (Groq, Cerebras) lie off-chart, trading capacity for bandwidth.}
\label{fig:designspace}
\end{figure}

Mapping the leading accelerators onto compute throughput versus memory bandwidth, as in Figure \ref{fig:designspace}, exposes the memory wall directly. Across NVIDIA's own line, peak BF16 compute grew roughly seven-fold from the A100 to the B200 while HBM bandwidth grew about four-fold, so the cluster of parts drifts steadily above any line of constant arithmetic intensity---compute is outrunning the bandwidth needed to feed it on memory-bound work. The HBM-based parts (the NVIDIA, AMD, Google, and AWS devices) form a band, with the newest training-class chips pushing both axes together. Two designs sit deliberately off this band and illustrate alternative bets: Groq's LPU forgoes HBM for hundreds of megabytes of on-chip SRAM at roughly 80 TB/s, and Cerebras's wafer-scale engine carries tens of gigabytes of SRAM at petabyte-per-second bandwidth, both purchasing bandwidth through SRAM at the cost of capacity and, ultimately, many chips per model. Processing-in-memory represents the opposite move---accepting modest compute to push effective bandwidth far higher (Figure \ref{fig:memhier}).

\subsection{Memory Hierarchy and Energy}
\begin{figure}[htbp]
\centering
\includegraphics[width=0.9\linewidth]{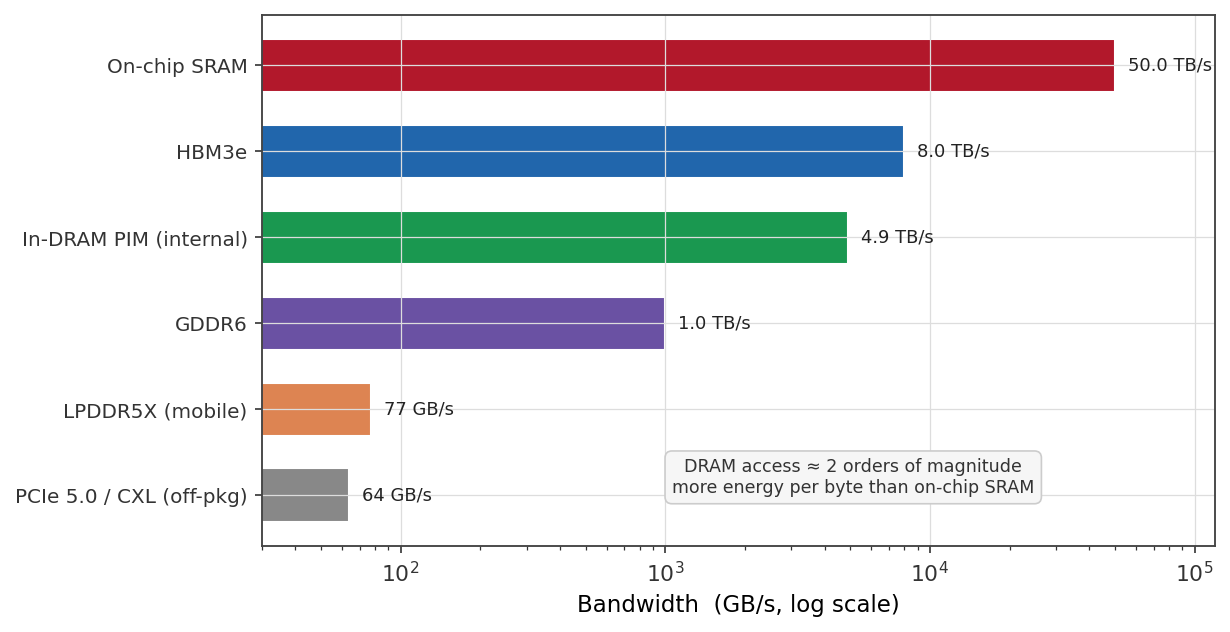}
\caption{Memory bandwidth hierarchy underlying the memory wall (log scale). A DRAM access costs roughly two orders of magnitude more energy per byte than an on-chip SRAM access.}
\label{fig:memhier}
\end{figure}

The bandwidth tiers underlying these choices span several orders of magnitude, as Figure \ref{fig:memhier} shows: on-chip SRAM at tens of terabytes per second, HBM3e at roughly five to eight, in-DRAM PIM at around five internally, conventional GDDR near one, and mobile LPDDR below a tenth. The energy dimension is equally steep and is the deeper reason the hierarchy exists: a DRAM access costs roughly two orders of magnitude more energy per byte than an on-chip access \cite{horowitz2014computing}, and bandwidth has scaled far more slowly than compute across generations \cite{gholami2024memorywall}. This single fact is the physical basis of both the memory wall that motivates processing-in-memory (Section 7) and the bandwidth ceiling that bounds on-device generation (Section 10); it is why proximity of compute to data, not raw arithmetic, increasingly governs both performance and efficiency.

\subsection{Energy Efficiency and the Limits of Nameplate Metrics}
Peak TOPS and TOPS-per-watt are nameplate figures, and they mislead for LLM serving in two ways: utilization is low on memory-bound decode, so realized throughput falls well short of peak, and end-to-end energy is dominated by data movement rather than the multiply-accumulates the spec sheet counts. Standardized benchmarks such as MLPerf Inference exist precisely because peak specifications do not predict delivered performance \cite{reddi2020mlperf}. The verified workload figures gathered across this review should be read in that light. Table \ref{tab:verified} consolidates them, and its caveat column is as informative as its results: the figures are not mutually comparable, since each reflects a different model, batch size, and operation. What survives the caveats is one consistent lesson: specialization yields large gains on the specific operation it targets, not uniform end-to-end speedups.

\begin{table}[htbp]
\centering
\caption{Verified workload-level results reported for non-GPU platforms in this review. Figures are not mutually comparable; each reflects a different model, batch size, and operation.}
\label{tab:verified}
\footnotesize
\setlength{\tabcolsep}{4pt}
\resizebox{\textwidth}{!}{
\begin{tabular}{lllll}
\toprule
Platform & Class & Workload & Reported result & Principal caveat \\
\midrule
Groq LPU (rack scale) & SRAM ASIC & Llama-2/3 70B decode & $\approx$300 tokens/s/user ($\approx$10\texttimes H100); 1--3 J/token & 576 chips across $\approx$9 racks (Section 5) \\
Samsung HBM2-PIM & In-DRAM PIM & GPT-1.3B & $\approx$348 tokens/s; $\approx$1.6\texttimes A100 & Memory-bound GEMV portions only (Section 7) \\
SK Hynix AiMX & In-DRAM PIM & OPT-6.7B, batch 1 & $\approx$330 tokens/s & Scale-out prototype (Section 7) \\
DFX & Multi-FPGA & GPT-2 1.5B & $\approx$120 tokens/s & 4\texttimes Alveo U280 appliance (Section 6) \\
FlightLLM & FPGA & LLaMA-2 7B, batch 1 & 6.0\texttimes energy-, 1.8\texttimes cost-efficiency vs.\ V100S; 1.2\texttimes A100 (VHK158) & Single-user, batch-1 regime (Section 6) \\
AttAcc & GPU + HBM-PIM & 175B-class batched decode & Up to 2.81\texttimes performance, 2.67\texttimes energy vs.\ iso-capacity GPU system & Simulation study (Section 7) \\
IBM NorthPole & Near-memory & Image recognition & Up to 5x H100 energy efficiency & Vision task, not language (Section 8) \\
SpikeGPT & Neuromorphic & 46M/216M language model & $\approx$5\texttimes lower energy & 2--3 orders of magnitude below frontier scale (Section 8) \\
\bottomrule
\end{tabular}}
\end{table}

\subsection{Historical Trajectory}
\begin{figure}[htbp]
\centering
\includegraphics[width=\linewidth]{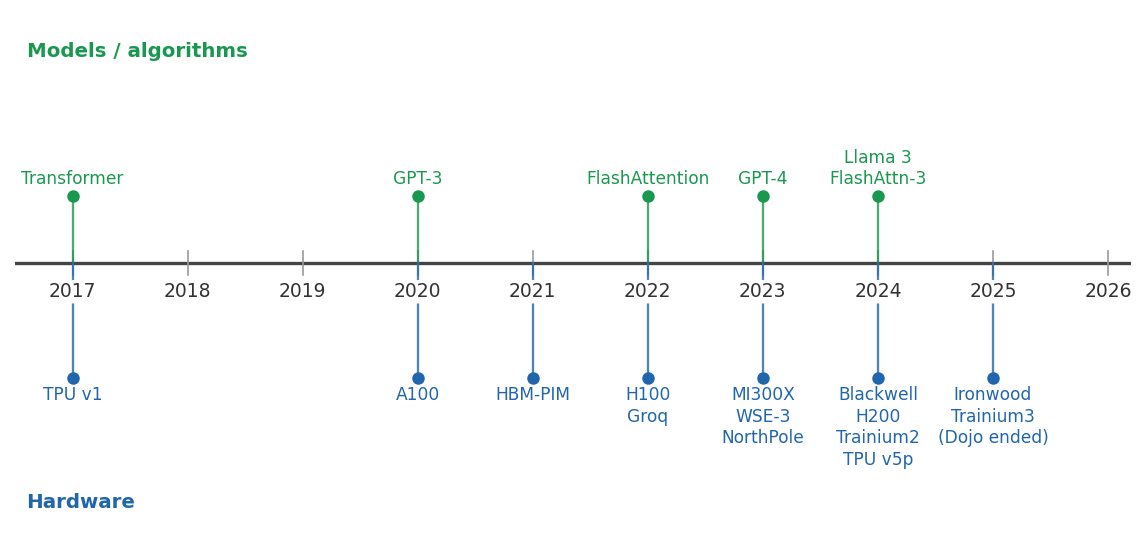}
\caption{Co-evolution of LLM models and algorithms (top) and accelerator hardware (bottom), 2017--2026.}
\label{fig:timeline}
\end{figure}

The timeline in Figure \ref{fig:timeline} traces two intertwined arcs from 2017 to 2026. The first is relentless scaling: the transformer and the first TPU in 2017, GPT-3 and the A100 in 2020, the H100 and FlashAttention in 2022, the Blackwell generation with Trainium2 and the H200 in 2024, and Ironwood and Trainium3 in 2025. The second is progressive specialization, from general-purpose GPUs toward domain-specific ASICs and, more recently, memory-centric designs. That specialization has casualties as well as successes, as Tesla's 2025 wind-down of the bespoke Dojo program illustrates (Section 7); the history is not one of uniform progress but of a widening, and occasionally pruned, set of architectural bets.

\subsection{Synthesis: No Single Winner}
Tables \ref{tab:qual} and \ref{tab:quant} consolidate the comparison qualitatively and quantitatively. The bottom line is that the field has no single winner and is unlikely to acquire one. GPUs remain the default for training and for flexible or lower-volume inference, on the strength of their software ecosystems and generality. Domain ASICs---TPUs, Trainium, Groq, and Cerebras---win at scale for stable, high-volume workloads where their efficiency justifies the loss of flexibility and the ecosystem commitment. FPGAs occupy a latency-sensitive and streaming niche. Processing-in-memory and near-memory computing are the most promising near-term response to the memory wall and are entering production as heterogeneous complements rather than replacements. Neuromorphic and photonic computing are not yet production platforms for LLMs, with photonics' near-term value lying in interconnect rather than computation. The durable pattern across all of this is heterogeneity: the systems that will serve large models combine these approaches rather than selecting one. Table \ref{tab:matrix} inverts the same conclusions, organizing them by deployment scenario rather than by platform class.

\begin{table}[htbp]
\centering
\caption{Qualitative comparison of accelerator classes for LLM workloads.}
\label{tab:qual}
\footnotesize
\renewcommand{\arraystretch}{1.25}
\begin{tabularx}{\linewidth}{l X X X X X}
\toprule
Class & Representative examples & Peak compute / memory model & Software maturity & Best-fit LLM role & Key limitation \\
\midrule
GPU & NVIDIA H100/B200, AMD MI300X & Very high; HBM, balanced & Very mature (CUDA/ROCm) & Training; flexible inference & Cost, power, HBM bandwidth ceiling \\
Domain ASIC & Google TPU, AWS Trainium, Groq, Cerebras & High--very high; HBM or large SRAM & Moderate, vendor-specific & High-volume training/inference at scale & Inflexibility; ecosystem lock-in \\
FPGA & AMD Alveo, Intel Agilex & Moderate; on-chip + limited HBM & Improving but effortful (HLS) & Low-latency, streaming niches & Bandwidth; toolchain effort \\
PIM / near-memory & Samsung HBM-PIM, SK Hynix AiM, CXL-PNM & Low compute, high bandwidth; compute in/near DRAM & Immature & Memory-bound attention/GEMV (as complement) & Limited compute; density; programmability \\
Neuromorphic & Intel Loihi 2, IBM NorthPole, BrainChip Akida & Low (event-driven); co-located & Immature (SNN tooling) & Edge / small or distilled models & Cannot run frontier-scale LLMs today \\
Photonic & Lightmatter, Celestial AI, research ONNs & High (linear ops only); no optical memory & Very immature & Interconnect (near-term); compute (long-term) & Precision; conversion overhead; no optical memory \\
\bottomrule
\end{tabularx}
\end{table}

\begin{table}[htbp]
\centering
\caption{Representative leading device per class (nameplate BF16/FP16, dense).}
\label{tab:quant}
\footnotesize
\setlength{\tabcolsep}{4pt}
\resizebox{\textwidth}{!}{
\begin{tabular}{lllllll}
\toprule
Class & Representative device & Peak compute & On-package memory & Bandwidth & Power & Maturity \\
\midrule
GPU & NVIDIA B200 & 2,250 TFLOPS & 192 GB HBM3e & 8 TB/s & \textasciitilde1,000 W & Production \\
Cloud ASIC (HBM) & Google TPU v7 ``Ironwood'' & 2,307 TFLOPS (4,614 FP8) & 192 GB HBM3e & 7.37 TB/s & \textasciitilde600 W & Production \\
Cloud ASIC (SRAM) & Groq LPU & 188 TFLOPS & 230 MB SRAM & \textasciitilde80 TB/s (on-chip) & \textasciitilde750 W class & Production (multi-chip) \\
Wafer-scale & Cerebras WSE-3 & 125 PFLOPS (sparse) & 44 GB SRAM & 21 PB/s (SRAM) & \textasciitilde23 kW (system) & Production \\
FPGA & AMD Alveo U280-class & tens of TFLOPS (INT8) & 8 GB HBM2 & \textasciitilde0.46 TB/s & \textasciitilde225 W & Production (niche) \\
PIM & Samsung HBM-PIM / SK Hynix AiM & per-bank MAC units & DRAM & \textasciitilde4.9 TB/s internal & --- & Prototype \\
Neuromorphic & Intel Hala Point (1,152\texttimes Loihi 2) & 1.15B neurons & on-chip & --- & \textasciitilde2.6 kW & Research \\
Photonic & Lightmatter Passage (interconnect) & --- & --- & 114 Tbps optical I/O & --- & Early product \\
\bottomrule
\end{tabular}}
\end{table}

\begin{table}[htbp]
\centering
\caption{Deployment-oriented recommendation matrix: the comparative conclusions of this section organized by workload scenario rather than by platform class.}
\label{tab:matrix}
\footnotesize
\setlength{\tabcolsep}{4pt}
\resizebox{\textwidth}{!}{
\begin{tabular}{llll}
\toprule
Workload scenario & Default today & Credible alternative & Emerging complement \\
\midrule
Frontier-scale training & GPU clusters (H100/B200 class) & TPU / Trainium pods & Co-packaged optics for scale-out \\
Flexible training and fine-tuning & GPU, single node to cluster & Cost-optimized cloud TPU tiers & --- \\
High-volume, stable cloud inference & TPU / Trainium / custom ASIC & GPU with vLLM-class serving & PIM offload of attention \\
Latency-critical interactive serving & SRAM-resident ASIC (Groq, Cerebras) & GPU + speculative decoding & --- \\
Long-context, memory-bound serving & Max-HBM GPU (H200/B200) + paged KV cache & CXL memory expansion & HBM-PIM / near-memory offload \\
Edge / on-device inference & Mobile NPU/SoC, $\leq$4B INT4 models & Embedded FPGA (streaming, deterministic) & Neuromorphic, milliwatt-class \\
\bottomrule
\end{tabular}}
\end{table}

\section{Open Challenges and Future Directions}
The preceding sections surface a consistent set of unresolved problems that cut across hardware classes. This section organizes them and closes with a practical selection guide (Figure \ref{fig:decision}).

\subsection{The Memory Wall Remains the Defining Constraint}
If one theme unifies this review, it is that memory---its bandwidth, capacity, and the energy of moving data---governs LLM acceleration more than arithmetic does. The decode phase is bandwidth-bound, the key-value cache can rival the model in size, and a DRAM access costs roughly two orders of magnitude more energy than an on-chip one. The responses now in motion are correspondingly memory-centric: continued high-bandwidth-memory scaling toward HBM4, the maturation of processing-in-memory and near-memory computing as heterogeneous complements (Section 7), CXL-based memory pooling and disaggregation, and software-level KV-cache management through paging, quantization, and compression. Two developments sharpen this picture. First, the constraint is increasingly a packaging problem as much as a memory one: HBM reaches the compute die only through advanced 2.5D integration, and interposer and stacking capacity---alongside the HBM supply shortage noted in Section 1---has repeatedly bounded accelerator production \cite{iea2026energy}. The HBM4 generation, which doubles the interface width to 2,048 bits and moves toward base dies customized to the host accelerator \cite{jedec2025hbm4}, blurs the memory--logic boundary further and turns memory vendors into co-design partners rather than commodity suppliers. Second, the workload itself is moving deeper into the constrained regime: reasoning models that expend test-time compute multiply the number of decode tokens generated per query \cite{snell2024testtime,deepseek2025r1}, shifting the serving mix further down the bandwidth-bound slope of the roofline and raising the economic weight of every memory-centric technique surveyed here. These are reinforced from the algorithm side by attention variants such as multi-query, grouped-query, and multi-head latent attention and by sparse and linear attention. The throughline is that future gains will come more from moving data less than from adding FLOPS.

\subsection{The Software and Programmability Gap}
The single largest practical barrier to adopting non-GPU hardware is software. The maturity of CUDA and its surrounding libraries is a durable moat, and every alternative examined here---domain ASICs, FPGAs, processing-in-memory, neuromorphic, and photonic---is gated less by raw capability than by the immaturity of its compilers, kernels, and toolchains. The most promising developments are portable compiler infrastructures and intermediate representations that decouple models from any single backend, including MLIR \cite{lattner2021mlir} and tile-based kernel languages such as Triton \cite{tillet2019triton}, alongside ecosystem efforts like OpenXLA and IREE. Until these abstractions mature, a hardware advantage on paper will not reliably translate into a deployed one, a pattern visible repeatedly in the gap between announced and delivered performance for novel architectures. Table \ref{tab:software} summarizes the programming model, key compiler infrastructure, and software maturity of each platform class.

\begin{table}[htbp]
\centering
\caption{Programming model, compiler infrastructure, and software maturity by accelerator class.}
\label{tab:software}
\footnotesize
\setlength{\tabcolsep}{5pt}
\renewcommand{\arraystretch}{1.2}
\begin{tabularx}{\textwidth}{l X X X l}
\toprule
Platform & Programming model & Key compiler / IR & Serving stack & Maturity \\
\midrule
GPU (NVIDIA) & CUDA, Triton & CUDA, TensorRT, MLIR & vLLM, TensorRT-LLM & Very mature \\
GPU (AMD) & HIP / ROCm & ROCm, MLIR & vLLM (partial) & Improving \\
TPU / Trainium & JAX, PyTorch/XLA & XLA, OpenXLA, IREE & JetStream, vLLM & Mature (vendor) \\
Groq / Cerebras & Vendor SDK & Proprietary graph compiler & Vendor runtime & Moderate \\
FPGA & HLS / RTL, FINN & Vitis HLS, FINN & Custom & Effortful \\
PIM / near-memory & Vendor library & Research / proprietary & Prototype only & Immature \\
Neuromorphic & Lava, SNN frameworks & SNN toolchains & Research only & Immature \\
Photonic & Research toolflows & Research & Research only & Very immature \\
\bottomrule
\end{tabularx}
\end{table}

\subsection{Energy Efficiency and Sustainability}
The energy trajectory of AI infrastructure, introduced in Section 1, makes efficiency a first-class design objective rather than a secondary virtue. Two needs follow. First, efficiency must be reported in delivered rather than nameplate terms, which requires standardized power-inclusive benchmarking of the kind MLPerf has begun to provide \cite{reddi2020mlperf}, because peak TOPS-per-watt does not predict the energy of a real serving workload. Second, since data movement dominates that energy, the architectural and algorithmic levers that reduce movement---low-precision formats, sparsity, and compute-near-data---are also the principal sustainability levers, aligning the efficiency and capability agendas rather than trading them off.

\subsection{Hardware--Algorithm Co-Design}
The most productive frontier is the co-evolution of models and hardware. Quantization to 4-bit and below, structured and mixture-of-experts sparsity, attention-mechanism redesign, and speculative decoding all advance in step with hardware support for native low-precision formats, sparse datapaths, and KV-aware memory hierarchies. The unsolved cases are instructive: a mixture-of-experts design that is practical within a mobile power and memory budget does not yet exist (Section 10), and reconciling extreme quantization with accuracy on the hardest models remains an active area. Progress here is increasingly a joint optimization over the model, its numerics, and the memory system, not over any one in isolation.

\subsection{Heterogeneity, Disaggregation, and Emerging Architectures}
\begin{figure}[htbp]
\centering
\includegraphics[width=\linewidth]{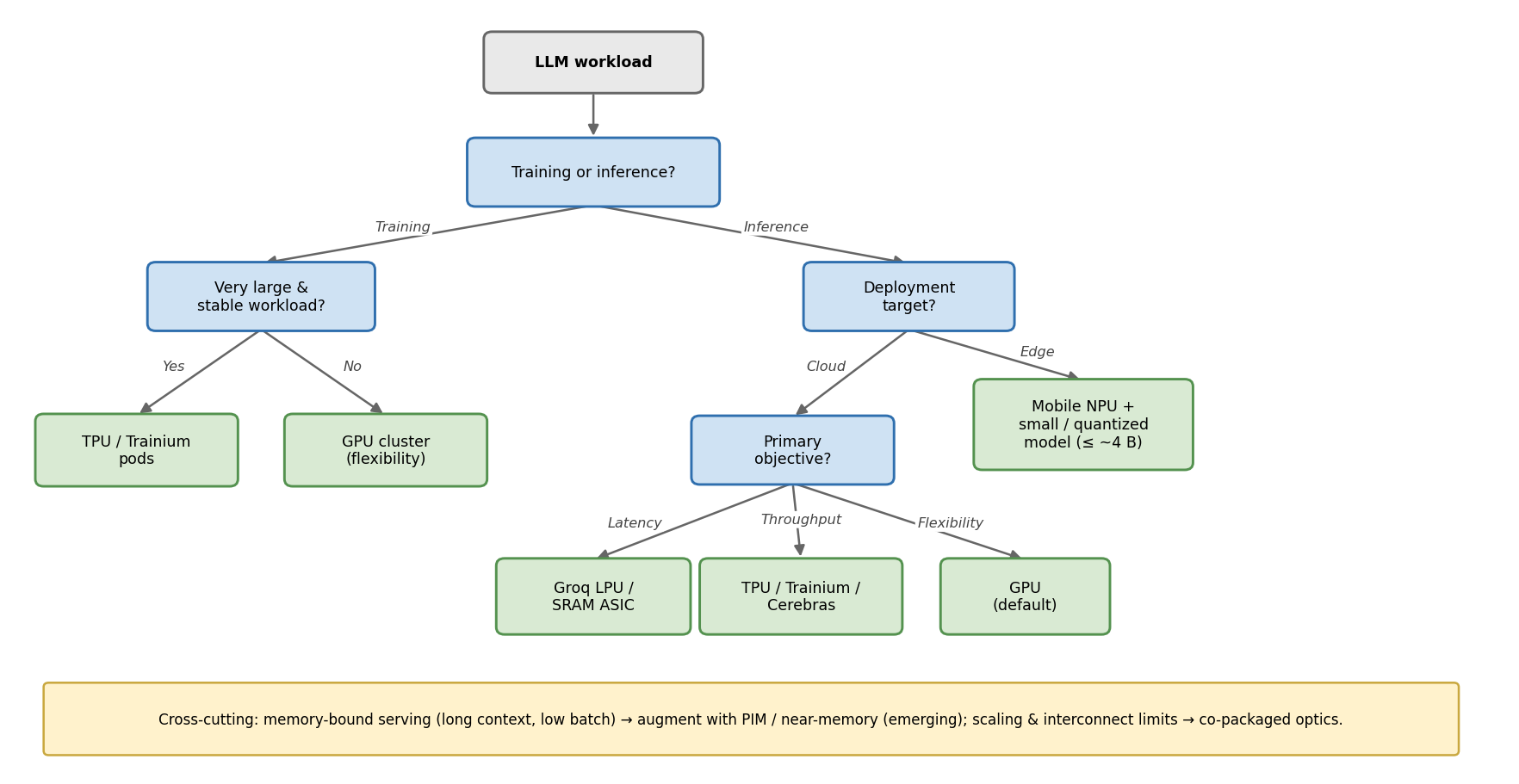}
\caption{A heuristic decision guide for selecting an accelerator platform from workload characteristics.}
\label{fig:decision}
\end{figure}

The comparative analysis of Section 11 pointed away from convergence on a single architecture and toward heterogeneous, disaggregated systems: compute-bound work on GPUs or NPUs, memory-bound attention on processing-in-memory, prefill and decode split across separate device pools at the serving layer \cite{patel2024splitwise,zhong2024distserve}, and memory pooled and scaled across CXL and optical fabrics. The emerging architectures enter this picture on a maturity gradient---processing-in-memory in the near term, analog in-memory computing and photonic interconnect in the medium term, and neuromorphic and photonic computation over a longer horizon---and it is important to be candid that the latter are not yet production platforms for LLMs. Figure \ref{fig:decision} distills how these considerations map onto a platform choice in practice.

\subsection{Benchmarking and Access}
Cutting through nameplate claims requires fair, workload-representative benchmarks and reproducible, power-inclusive reporting; the community's benchmarking efforts are necessary infrastructure rather than an afterthought \cite{reddi2020mlperf}. A related concern is access: the most capable accelerators are concentrated among a few vendors and well-resourced organizations, and cost and supply constraints shape who can train and serve large models. Open-weight models, more efficient inference, and increasingly open hardware and toolchains broaden participation, but the gap is real and worth tracking as the field matures.

\subsection{Outlook}
The binding constraints on LLM acceleration are the memory wall and the software gap, not a shortage of arithmetic. The most credible path forward runs through hardware--algorithm co-design and heterogeneous, memory-centric systems, with emerging architectures adopted selectively where their strengths align with a specific operation or deployment. No single platform wins outright, and the field is better understood as assembling a toolkit than as searching for a successor to the GPU---a view the conclusion develops.

\section{Conclusion}
This review surveyed the landscape of AI hardware accelerators for large language models, spanning general-purpose GPUs, custom ASICs, reconfigurable FPGAs, processing-in-memory and near-memory designs, and the emerging neuromorphic and photonic approaches, across both cloud and edge deployment. The motivating tension throughout has been the collision between the explosive growth of LLMs and the physical and economic limits of the hardware that must train and serve them. Rather than rank platforms, we organized them by architecture, deployment target, and optimization objective, and grounded the comparison in verified specifications and a roofline-based account of what the workload actually demands.

The central finding is that memory, not arithmetic, is the defining constraint on LLM acceleration. The decode phase that dominates serving---increasingly so as reasoning models multiply the tokens generated per query---is bandwidth-bound, the key-value cache can rival the model weights in size, and the energy of moving data exceeds that of computing on it by orders of magnitude. This reframes acceleration as a problem of moving data less rather than computing faster, and it explains why the most active responses are memory-centric: continued high-bandwidth-memory scaling, the maturation of processing-in-memory as a heterogeneous complement, and memory pooling over CXL and optical fabrics. The roofline and bandwidth-hierarchy analyses of Section 11 show this constraint operating identically in the datacenter and on the phone.

It follows that the field has no single winner and is better understood as a toolkit than as a search for a successor to the GPU. GPUs remain the flexible default and the workhorse of training, on the strength of their software ecosystems; domain-specific ASICs such as TPUs, Trainium, Groq, and Cerebras win at scale for stable, high-volume workloads where their efficiency justifies the loss of generality; FPGAs hold a latency-sensitive and streaming niche; and processing-in-memory is the most promising near-term answer to the memory wall, entering production as a complement rather than a replacement. Neuromorphic and photonic computing are genuinely promising but are not yet production platforms for frontier-scale LLMs, with photonics' nearest-term value lying in interconnect rather than computation. The honest characterization is differentiation by workload, captured in the decision guide of Section 12.

The binding constraints going forward are therefore the memory wall and the software and programmability gap, not a shortage of compute. The maturity of CUDA remains a decisive advantage, and every alternative architecture is gated less by its raw capability than by the immaturity of its compilers and toolchains; portable compiler infrastructures are as important to the future of this field as any silicon. The most credible path runs through hardware--algorithm co-design---quantization, sparsity, mixture-of-experts, and attention redesign co-evolving with hardware---and through heterogeneous, disaggregated, memory-centric systems. Energy efficiency reported in delivered rather than nameplate terms, and fair power-inclusive benchmarking, are cross-cutting imperatives that bear on both the sustainability and the accessibility of large-scale AI.

In sum, the decisive advances in LLM acceleration are unlikely to come from any one architecture displacing the others. They will come from assembling a heterogeneous toolkit matched to the diversity of LLM workloads, and from optimizing jointly across the model, its numerics, and the memory system. The accelerator landscape is converging not on a single design but on a shared recognition that, for large language models, the memory system is the computer.

\section*{Declaration of competing interest}
The authors declare that they have no known competing financial interests or personal relationships that could have appeared to influence the work reported in this paper.


\bibliographystyle{unsrt}
\bibliography{references}

\end{document}